\documentclass[aps,twocolumn,superscriptaddress,prb]{revtex4-2}
\usepackage[utf8]{inputenc}
\usepackage{amsmath, amssymb, graphicx, color, subfigure, enumitem}
\usepackage[bookmarks=false]{hyperref}
\usepackage[usenames,dvipsnames]{xcolor}
\usepackage[super]{nth}
\usepackage[normalem]{ulem}

\def\be{\begin{equation}}
\def\ee{\end{equation}}
\def\ba{\begin{eqnarray}}
\def\ea{\end{eqnarray}}

\begin{document}

\title{Gapless superconductivity in the low-frequency electrodynamic response of two-dimensional granular In/InO$_x$ composites}

\author{Xinyang Zhang}
\affiliation{Geballe Laboratory for Advanced Materials, Stanford University, Stanford, CA 94305, USA}
\affiliation{Department of Applied Physics, Stanford University, Stanford, CA 94305, USA}

\author{Jinze Wu}
\affiliation{Geballe Laboratory for Advanced Materials, Stanford University, Stanford, CA 94305, USA}
\affiliation{Department of Physics, Stanford University, Stanford, CA 94305, USA}

\author{Alexander Palevski}
\affiliation{School of Physics and Astronomy, Raymond and Beverly Sackler Faculty of Exact Sciences, Tel Aviv University, Tel Aviv 6997801, Israel}

\author{Aharon Kapitulnik}
\affiliation{Geballe Laboratory for Advanced Materials, Stanford University, Stanford, CA 94305, USA}
\affiliation{Department of Applied Physics, Stanford University, Stanford, CA 94305, USA}
\affiliation{Department of Physics, Stanford University, Stanford, CA 94305, USA}

\date{\today}

\begin{abstract}
We measured the full complex ac conductance of two-dimensional granular In/InO$_x$ composites using the mutual inductance technique to explore the transition from a ``failed-superconductor-turned anomalous metal’’ to a robust superconductor. In this system, room-temperature annealing was adopted to tune the InO$_x$-mediated coupling between In grains, allowing for the observation of both a ``true’’ superconductor-to-insulator transition and the emergence of an intervening anomalous metallic state. In this paper, we show that further annealing increases the inter-grain coupling, which eliminates the anomalous metallic phase, but at the same time prevent the emergence of strong Bose-dominated insulating phase. The complex ac conductance revealed a $T\to0$ saturating dissipative response in a finite magnetic field, coexisting with a robust superfluid density. The anomalous power-law spectra for the dissipative response appear to indicate quantum critical behavior proximate to a quantum superconductor to anomalous-metal transition as probed in the kilo-Hertz range, and point to signatures of gapless superconductivity in our granular superconducting system.
\end{abstract}

\maketitle

\section{Introduction}

A two-dimensional (2d) granular superconductor with established intra-grain pair amplitude and inter-grain coupling can be described by a model system of random array of Josephson junctions \cite{abeles_effect_1977,efetov_phase_1980,simanek_instability_1980}. The ground states of such a system are determined by a competition of Josephson coupling and electrostatic charging \cite{abeles_effect_1977}, along with the potential influence of dissipative coupling to the environment, often through local shunt resistors \cite{fisher_quantum_1986,chakravarty_effect_1987}. While the conventional picture of $T\to 0$ phases of such an array has been developed based on the assumption that the only stable phases of any disordered two-dimensional electron fluid are superconducting or insulating, thus leading to extensive investigations of a putative  superconductor-insulator transition (SIT) \cite{fisher_boson_1989,beloborodov_granular_2007}, increasing evidence for the emergence of intervening anomalous metallic (AM) phases \cite{kapitulnik_colloquium_2019} challenges the conventional perception.

Previously, experimental studies of granular ultra-thin films \cite{orr_local_1985,jaeger_threshold_1986,orr_global_1986,liu_resistive_1992}, metal-alloy/oxide composites \cite{deutscher_percolation_1980,fiory_superconducting_1983,hebard_pair-breaking_1984}, artificial granular composites \cite{eley_approaching_2012,han_collapse_2014,bottcher_superconducting_2018,yang_intermediate_2019}, and certain amorphous thin films \cite{kowal_disorder_1994,okuma_field-induced_1995} where induced granularity may dominate the superconducting state \cite{ghosal_role_1998}, have uncovered a plethora of anomalies. These include quasi-reentrant resistivity \cite{orr_local_1985}, logarithmic \cite{steiner_possible_2005} or exponential \cite{liu_resistive_1992,merchant_crossover_2001,allain_electrical_2012} temperature dependence of resistivity, and enhanced/suppressed quantum fluctuations \cite{fisher_dissipation_1987}. In a 2d granular system of indium--indium-oxide (In/InO$_x$), where the amorphous InO$_x$ mediates the coupling between superconducting In grains, annealing was shown \cite{zhang_robust_2021} to control the oxygen-vacancy that eventually tunes the $T=0$ quantum phase transition from a direct SIT \cite{hen_superconductorinsulator_2021} to one that exhibits a cascade of Bose-dominated (i.e. Cooper pairs) quantum phase transitions from a superconductor to AM to insulator \cite{zhang_anomalous_2022}. At the same time, it is evident that with strong enough inter-grain coupling such as a continuous path of strongly connected In grains, a robust superconductor exhibiting a gap of order of that of pure In prevails, which upon an application of a magnetic field, will transition into a Fermi-dominated weakly localized metal displaying superconducting fluctuations \cite{Larkin1980}. Indeed, such a change in behavior is achieved with further annealing of our In/InO$_x$ system. Thus, questions arise about the transmutation from the Bose-dominated to the Fermi-dominated system, the manner by which the anomalous metal fades into a superconductor to weak-insulator (WI) transition, and in particular the nature of the superconducting state in this intermediate regime.

In this paper, we explore the crossover from a S-AM-I transitions sequence to a S-WI behavior, focusing on the properties of the superconducting state just as the anomalous metallic phase disappears. Our results and conclusions originate from a systematic study of the complex ac conductance of the 2d In/InO$_x$ granular system as a function of temperature, magnetic field, and frequency, complemented by dc resistance measurements. As the system was tuned by annealing such that the coupling among In was greatly enhanced, a superconducting ground state with anomalous ac response emerges. Here, superfluid density and dissipative response, which saturates as $T\to 0$, coexist in the zero-dc-resistance state. Frequency spectra for the dissipative response and calculated ac resistance follow power-laws, in distinct contrast to fully gapped superconductors. Together with the dc resistance, which on its own can be characterized by a finite temperature regime of scaling, reminiscent of previously measured transitions in low-$R_\Box$ MoGe films \cite{Mason1999}, the calculated ac resistance extends the measurement of the superconducting transition to $\sim 10^{-5}\ \Omega$, shedding light on the anomalous finite-frequency electrodynamic response of a potential gapless superconducting state.

Gapless superconductors do not develop a full energy gap in the quasi-particle excitation spectrum, yet the diamagnetic supercurrent screening is preserved \cite{de_gennes_superconductivity_1966}. Examples include s-wave superconductors with magnetic impurities, dirty superconductors in strong magnetic field, a proximitized superconductor-metal interface \cite{tinkham_introduction_1996}, and more akin to our system, thin films of varying thickness in a field \cite{de_gennes_superconductivity_1966}. In these systems, the time-reversal symmetry of the paired states are broken by pair-breaking perturbations, and Anderson's theorem ceases to apply, leading to lower $T_c$ and gapless superconductivity even at $T=0$. In systems exhibiting QSMT, the anomalous metal phase is presumably gapless \cite{kapitulnik_colloquium_2019}, whereas in an array of RSJ junctions coupled to a dissipative heat-bath, the region with large Josephson coupling yet weak dissipation for quantum fluctuations may exhibit an exotic ordered superconducting phase with finite conductivity \cite{fisher_dissipation_1987}. Finally, possible observation of gapless excitations with increasing granularity was deduced from the temperature dependence of the magnetic penetration depth in NbN films and was interpreted as being a result of thermally-excited inter-grain phase fluctuations  \cite{lamura_granularity-induced_2002}.

Evidence for gapless superconductor can be unambiguously deduced from the frequency dependence of its electrodynamic response, thus being the focus of the present paper. The inductive (imaginary) quadrature is related to the superfluid density or the penetration depth, which would exhibit non-BCS temperature dependence, whereas the dissipative (real) quadrature directly measures the dynamics of any gapless excitations \cite{tinkham_introduction_1996}. At low frequencies in the kilo-Hertz range, the mutual inductance (MI) technique has been employed to study the BKT transition in disordered thin films \cite{yazdani_observation_1993,yazdani_phase_1995} and weakly-disordered amorphous MoGe and InO$_x$ \cite{misra_measurements_2013}, where a discrepancy between the transport and the superfluid density quantum critical point alluded an intervening incoherent metallic phase. At high frequencies in the microwave regime, the complex conductivity was probed by microwave techniques \cite{crane_survival_2007,liu_microwave_2013}. In weakly-disordered InO$_x$, the data were in agreement with the low-frequency MI experiment \cite{misra_measurements_2013}, indicating a quantum superconductor-metal transition marked by the quenched superfluid density at the critical point. The electrodynamic response in this metallic state was highly anomalous \cite{breznay_particle-hole_2017,wang_absence_2018}.

\section{Methods and Material}

The complex ac conductance was measured using a precision-built high-sensitivity gradiometer-type MI probe as shown in Fig.~\ref{fig3}(a), pioneered by Ref.~\cite{jeanneret_inductive_1989}. The probe consists of a series of two astatically-wound receive coils, surrounded by a drive coil that is centro-symmetrical to the receive coils. Both coils were wound using fine insulated copper wires \cite{supplemental}. 

With the probe placed immediately above a thin film sample, an ac current was supplied to the drive coil, creating a minuscule ac magnetic field perturbation ($\sim$ mOe) in the sample. The inductive (screening) response of supercurrent and dissipative response of any quasiparticles in the system in turn induced a complex voltage $\tilde{V}\equiv V_R-iV_I$ across the lower receive coil, where the direct MI between the coils had been suppressed by the astatic-winding. Finally, the induced voltage was measured using a lock-in amplifier phase-locked to the drive ac current, whereas any temperature/field-independent background and circuit phase-shift were subsequently removed from the signal \cite{supplemental}.

Under an infinite-plane approximation for a thin film, which generally holds when the film's lateral size spans several times the receive coil diameter \cite{turneaure_numerical_1998}, this 2d electrodynamics problem can be solved using the Fourier method \cite{pearl_current_1964}. Given a precise knowledge of the coil geometry, through a numerical look-up procedure, one can extract the coil-geometry-independent complex ac conductance, $\tilde{G}\equiv G_R-iG_I$, from the measured complex voltage $\tilde{V}$ at frequency $\omega$ \cite{jeanneret_inductive_1989}. $G_R$ and $\omega G_I$ together constitute the complex response function of electrical current to a small electromagnetic field perturbation. Additionally, as the complex conductance measurement is more sensitive to low-level dissipation, it can be converted to $R=\text{Re}[\tilde{G}^{-1}]$ as a sensitive ac resistance probe in the $\mu\Omega\text{-m}\Omega$ range, complementary to standard current-biased four-point measurements.

The 2d granular In/InO$_x$ samples for MI measurements were within the same batch as those previously extensively studied \cite{hen_superconductorinsulator_2021,zhang_robust_2021,zhang_anomalous_2022}. The only differences are that these samples were grown on borosilicate glass substrates and had been further annealed at room temperature in nitrogen environment, which rendered increasingly conductive samples. We systematically studied their ac response in a dilution refrigerator using drive frequencies between 2 kHz and 100 kHz over the entire magnetic-field-temperature phase diagram. Electrical resistance was measured on a similar sample in a hall bar geometry using the standard four-point lock-in technique (1-nA excitation current at 23 Hz).

\section{Results and Discussions}

\subsection{Resistance measurements: superconductor-weak-insulator transition}

\begin{figure}[t]
	\centering
	\begin{subfigure}{}
		\includegraphics[width=\columnwidth]{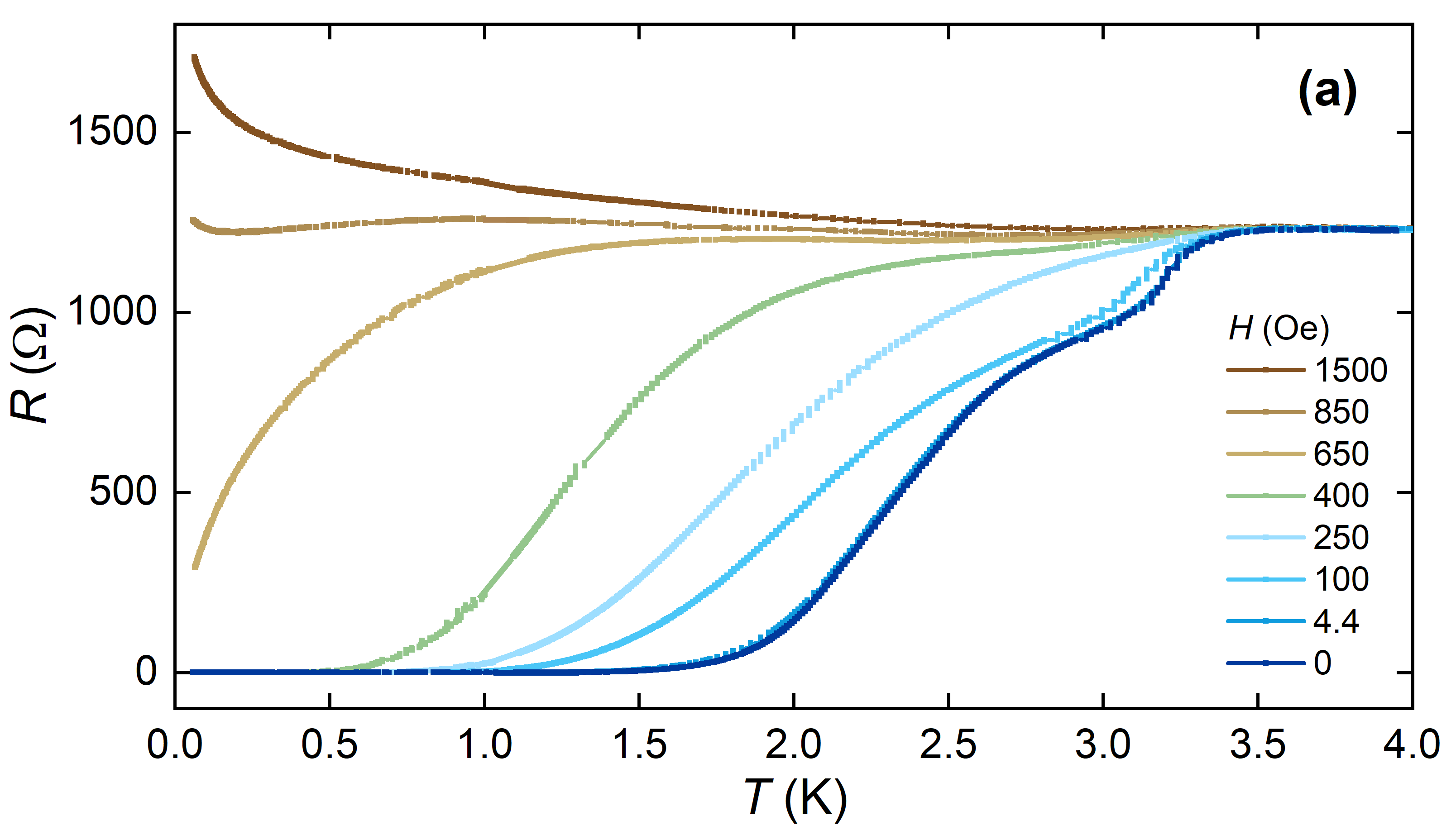}
	\end{subfigure}
	\vspace{-5mm}
	\begin{subfigure}{}
		\includegraphics[width=\columnwidth]{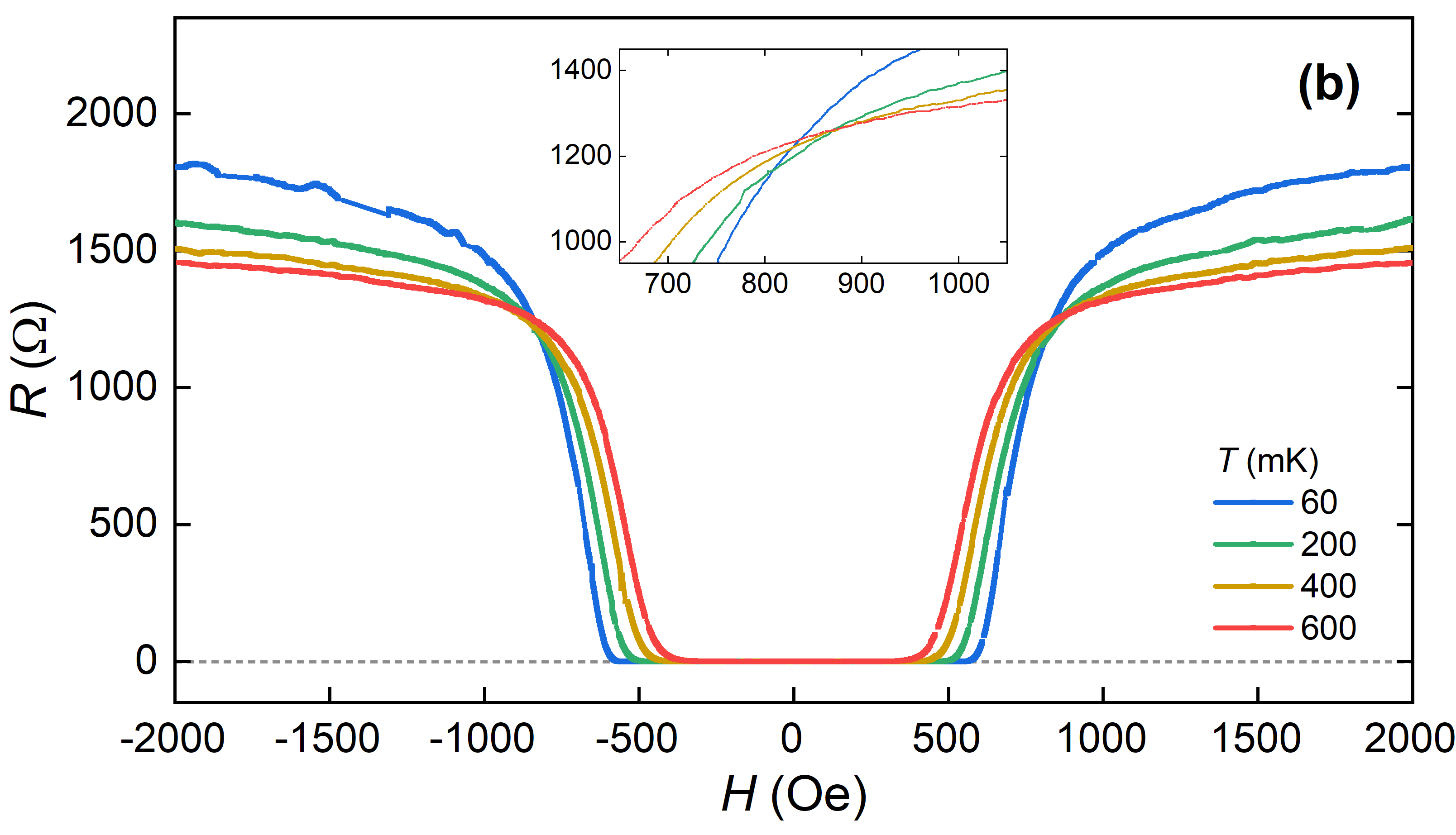}
	\end{subfigure}
	\caption{Resistance measurement. \textbf{(a)} Temperature- and \textbf{(b)} magnetic-field dependence of resistance were measured using the current-biased four-point method at 23 Hz. Inset in (b) shows the crossing point typically associated with a superconductor-insulator transition, as visually suggested by the transport data.}
	\label{fig1}
\end{figure}

\begin{figure*}[t]
	\centering
	\includegraphics[width=\textwidth]{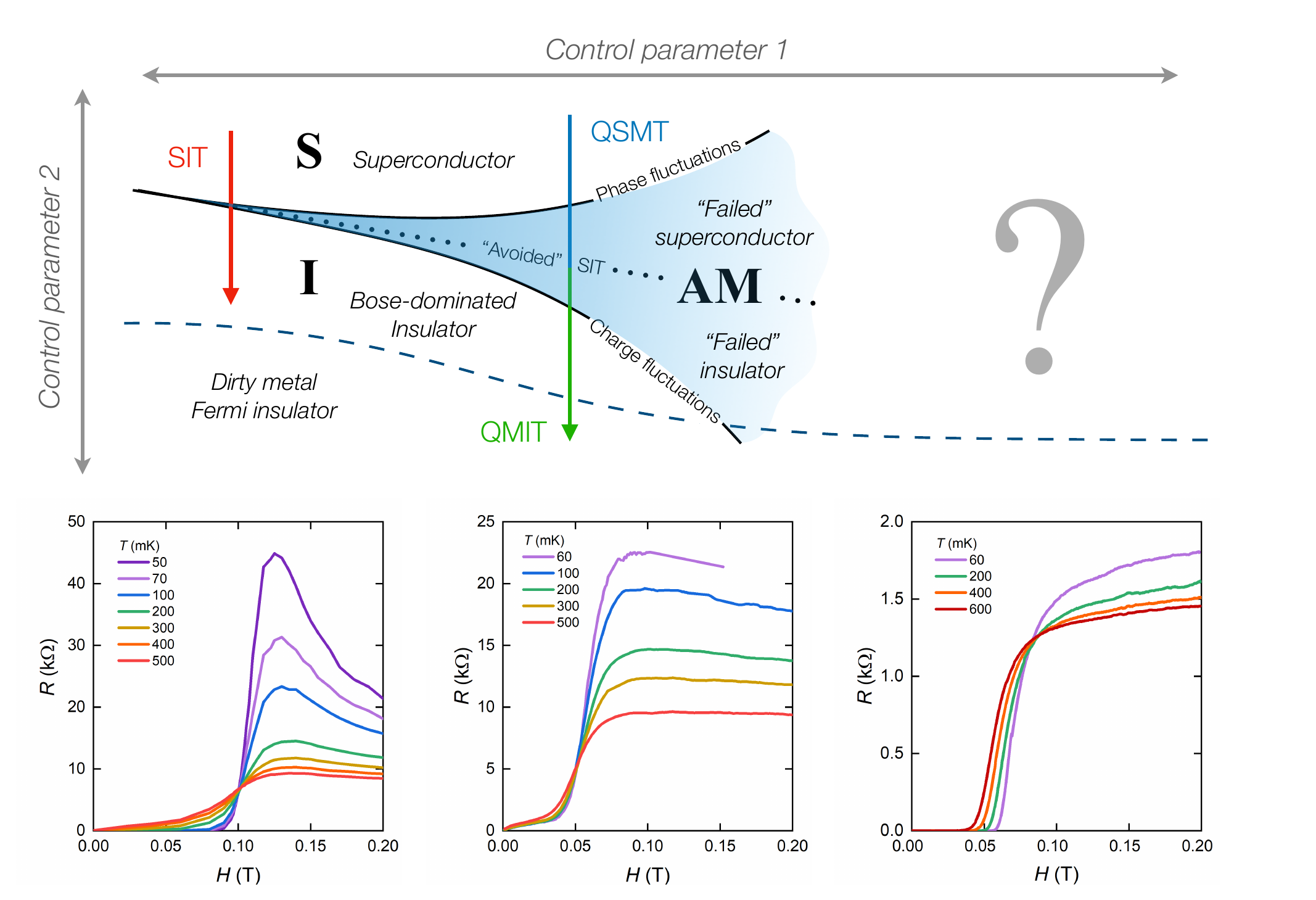}
	\caption{Zero-temperature phase diagram of granular In/InO$_x$ composite tuned by two control parameters (CPs). In the present study, CP 1 is annealing and CP 2 is magnetic field. The arrows represent, from the left to right, a direct SIT (red), a QSMT(blue)-QMIT(green), and the transition probed in this study. The magneto-resistance plots below were adapted respectively from Ref.~\cite{hen_superconductorinsulator_2021}, Ref.~\cite{zhang_robust_2021}, and Fig.~\ref{fig1}(b). The region marked by the question mark is central to this study.}
	\label{fig2}
\end{figure*}

The temperature and magnetic-field dependence of resistance of the further annealed sample are shown in Fig.~\ref{fig1}. Below the superconducting transition at $\sim 3.4$ K for In grains, the $T=0$ quantum phase transition seems to indicate a superconductor-weak-insulator transition. The resistance saturation associated with anomalous metallic phases observed in less-annealed samples \cite{zhang_robust_2021,zhang_anomalous_2022} is absent. The weak positive magnetoresistance (MR) exhibits a crossing point at $H^*\approx 850$ Oe between the isotherms. Such weak magnetic-field dependence has been pointing to anomalous metallic behavior in amorphous MoGe \cite{mason_true_2001} and InO$_x$ \cite{breznay_particle-hole_2017}, whereas a strong positive MR peak usually indicates a direct SIT \cite{sambandamurthy_superconductivity-related_2004,steiner_approach_2008}. Having measured resistivity data both with 
\cite{zhang_robust_2021} and without (this study) \textit{apparent} resistance saturation using exactly the same experimental set-up, we have demonstrated yet another evidence for the absence of spurious heating \cite{tamir_sensitivity_2019} in our experiments.

\subsection{Ground states of granular In/InO$_x$ tuned by annealing}

Composed of granular In deposited on a uniform underlying non-superconducting InO$_x$, the In/InO$_x$ system has been modeled as a granular superconductor with Josephson coupling mediated by the InO$_x$ \cite{zhang_anomalous_2022}, where the resistivity evolution as a function of magnetic field has been systematically explored across a series of annealing stages. Annealing at room temperature in vacuum was well known \cite{kowal_disorder_1994} to increase the oxygen vacancy in the electron-doped InO$_x$, yielding more conductive samples. 

In Fig.~\ref{fig2}, from left to right, we examine the MR and the $T\to0$ ground states of In/InO$_x$ system as tuned by annealing. Less-annealed superconducting samples \cite{hen_superconductorinsulator_2021} typically display a paradigmatic SIT at the quantum critical point $H^*$, as well as a large MR peak associated with the bose-dominated insulating phase before the superconducting order parameter diminishes. The critical field $H^*$ was found to be much smaller than the upper critical field $H_{c2}$ in amorphous InO$_x$ \cite{steiner_approach_2008}. In annealed samples, however, anomalous metallic phases often emerge as the MR peak appears less prominent, while an intervening metallic ground state conceals an ``avoided'' SIT at $H^*$ \cite{zhang_anomalous_2022}. The critical field $H^*$ usually shifts towards $H_{c2}$, or in granular superconductors, the single-grain critical field $H_{c}$.

Whatever the ground states are for any further annealed samples, as the annealing removes oxygen from InO$_x$, eventually we would arrive at an In film where the thickness fluctuates subject to the geometry of the deposited In grain. At $T=0$ we should expect a superconducting-normal transition, under the influence of fluctuating pair amplitude between the grains. Here, as shown in Fig.~\ref{fig1}(b), the critical field $H^*$ is close to the effective critical field of In: $H_{c,\text{eff}}\sim1000$ Oe, indicating pairing breaking effect and fermionic excitations at the superconductor-weak-insulator transition. In the following sections, we will examine the resultant superconducting ground state by probing the ac electrodynamic response of a further-annealed In/InO$_x$ sample in the zero-resistance regime.

\subsection{Complex ac conductance measurements: superconducting transitions as a function of frequency and magnetic field}

\begin{figure*}[t]
	\centering
	\includegraphics[width=\textwidth]{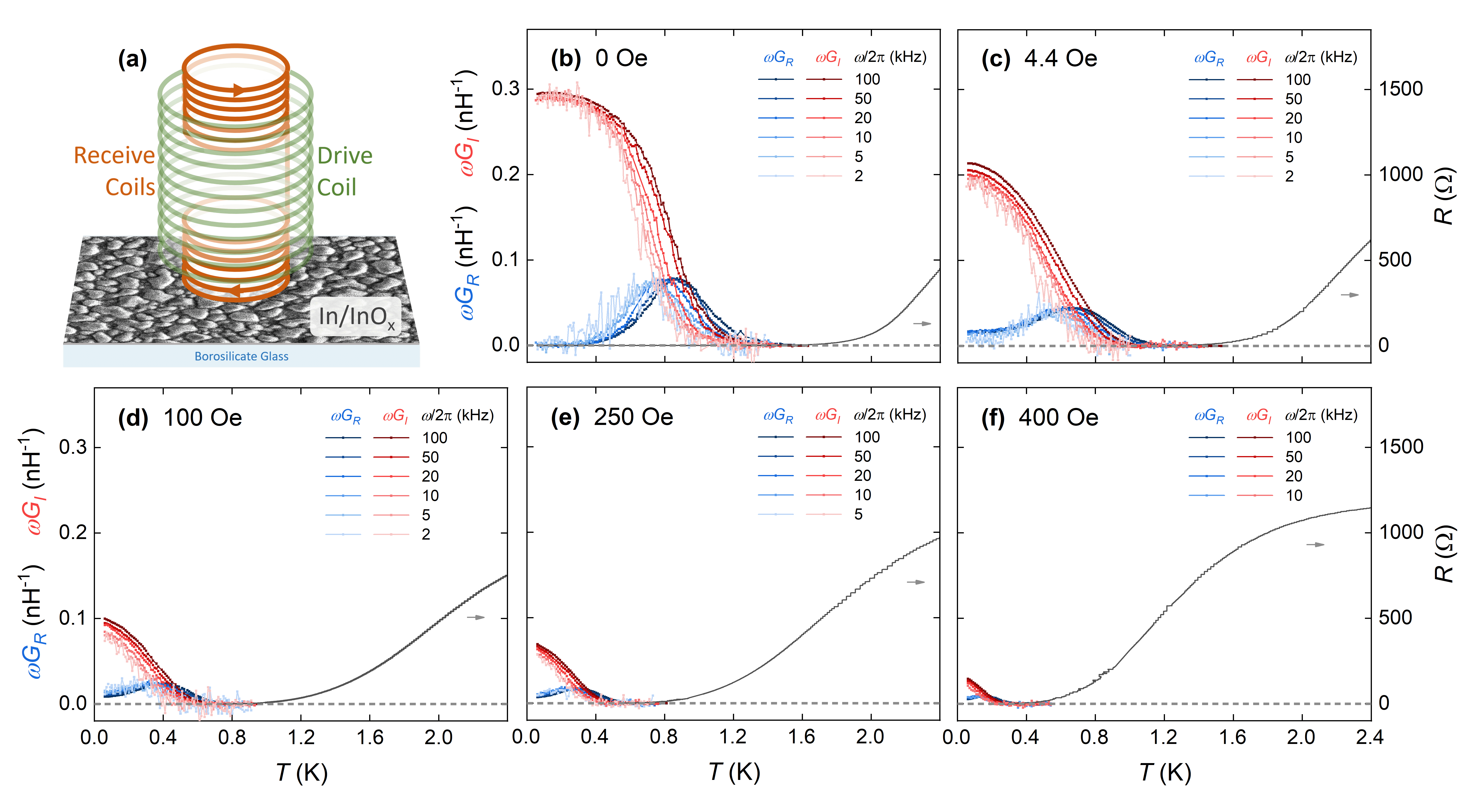}
	\caption{Temperature dependence of complex ac conductance at different frequencies and magnetic field. \textbf{(a)} Schematic diagram of the precision-built mutual inductance (MI) probe on a 2d granular In/InO$_x$ sample. \textbf{(b-f)} Real and imaginary part of the complex ac conductance $\omega G_R$ and $\omega G_I$ (left-axis) in different perpendicular magnetic field. Right axes correspond to resistance measurement in Fig. 1. Note the coexistence of superfluid response and finite dissipation in the presence of a magnetic field.}
	\label{fig3}
\end{figure*}

In Fig.~\ref{fig3}, we present a systematic measurement of the temperature dependence of complex ac conductance as a function of frequency and magnetic field. In zero magnetic field (Fig.~\ref{fig3}(b)), our 2d granular system undergoes a Berezinskii-Kosterlitz-Thouless (BKT) transition at $T_\text{BKT}\approx 1.3$ K, where the supercurrent screening response $\omega G_I$ increases from zero and saturates at the $T=0$ value following a broad superconducting transition \cite{yazdani_phase_1995,misra_measurements_2013}. The maximum value, or $T=0$ superfluid density, $\omega G_I(0)\propto\lambda^{-2}(0)$ is equivalent to a 2d Pearl length \cite{pearl_current_1964} of $\lambda_\perp\equiv \lambda^2/d\approx 2.7$ mm and an effective penetration depth $\lambda\approx 15\ \mu$m using a nominal thickness of $d=80$ nm. A jump in the superfluid density \cite{beasley_possibility_1979}, however, is not visible, as the transition is greatly smeared out by the broad granular distribution in our system \cite{zhang_anomalous_2022}. This broadening arises as MI measures an averaged response over a macroscopic sample region set by the coil diameter. For higher frequencies, the BKT transition temperature $T_\text{BKT}$ increases, consistent with the BKT dynamical theory \cite{ambegaokar_dynamics_1980}. Furthermore, in zero field, the imaginary part of response function $\omega G_I$ is frequency-independent as expected \cite{tinkham_introduction_1996}.

In a finite magnetic field (Fig.~\ref{fig3}(c-f)), both the transition temperature $T_\text{BKT}$ and the magnitude of the inductive response $\omega G_I$ are increasingly suppressed. At merely 4.4 Oe, $\omega G_I$ is already reduced by $\sim30\%$, whereas the dissipative response $\omega G_R$ saturates at a non-zero value as $T\to0$. This trend persists for stronger magnetic fields, until $\sim$500 Oe, when the superfluid response has largely vanished and a large number of superconducting In grains are driven normal. 

The finite dissipative response at finite field, coexisting with finite superfluid inductive response, is anomalous. In a fully gapped superconductor, the real/dissipative part of ac conductance $G_R$ would be exponentially suppressed as $T\to0$. Despite that the resistance measurement had suggested a zero-resistance superconducting ground state $\lesssim 500$ Oe, we observe finite temperature-independent dissipation throughout this regime, in the presence of finite superconducting correlation.

Both the real and imaginary response functions acquire frequency dependence in a finite magnetic field. The $T=0$ inductive response is larger for higher frequencies, which correspond to a shorter dynamical length scale $L_\omega\sim\sqrt{D/\omega}$ where $D$ is the diffusion constant. This indicates a granular effect on the supercurrent screening of the drive ac magnetic field, which will be discussed later in Secion III.E. It is important to note that, at each frequency, the same transmission line phase shift was reversed for all zero/finite-field data, such that the dissipative response $\omega G_R=0$ at $H=0$ as $T\to0$, which is our main assumption as we expect a zero-field superconducting ground state. See \textit{Supplemental Material} \cite{supplemental} for details of data processing.

\subsection{Magnetic-field dependence of the complex ac conductance}

\begin{figure}[t]
	\centering
	\includegraphics[width=\columnwidth]{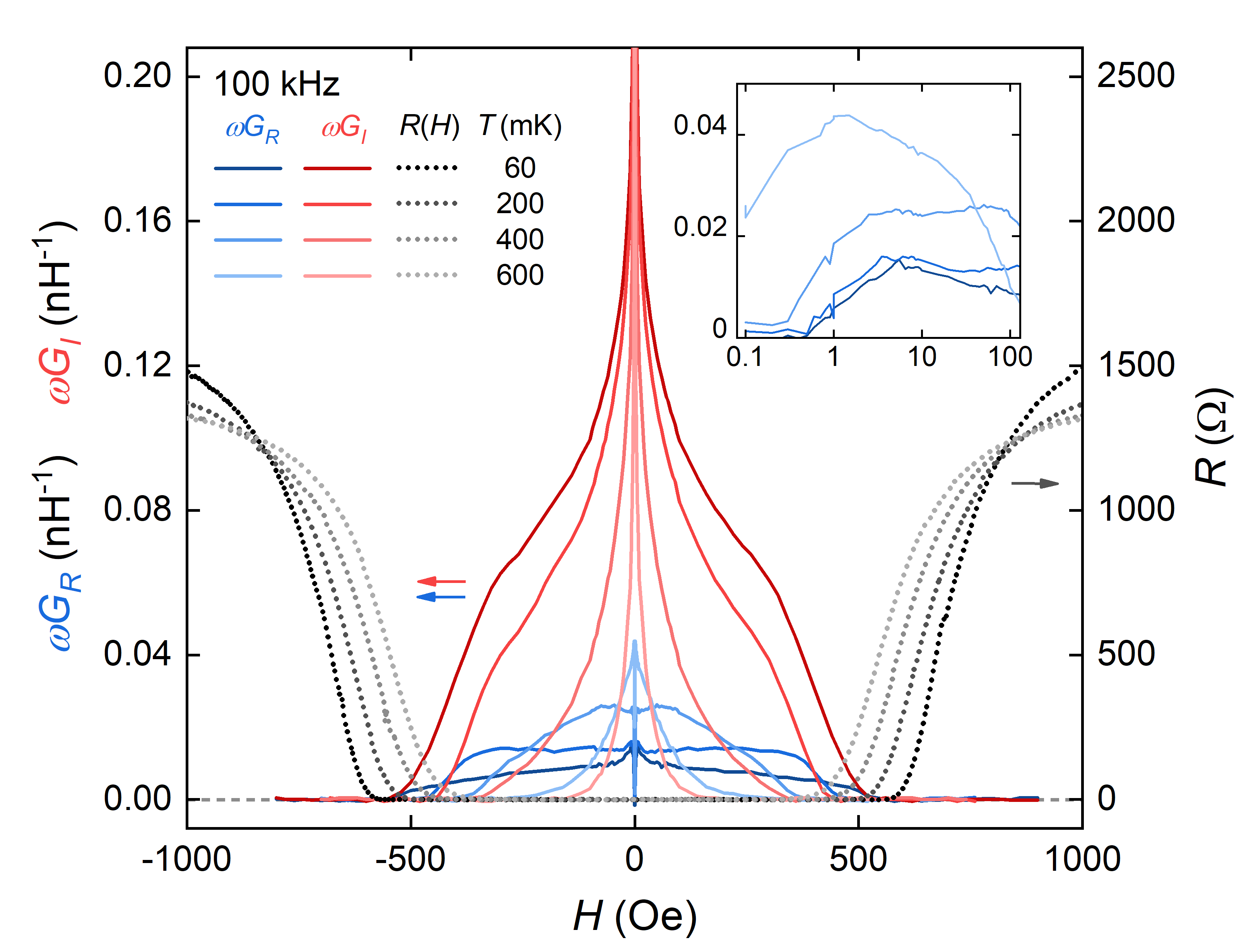}
	\caption{Magnetic-field dependence of complex ac conductance at 100 kHz at different temperatures. Magnetoresistance from four-point resistance measurement are also included (right-axis). Self-consistent with the temperature-dependence data, this plot highlights the sharp onset of the dissipative response near $H=0$. Inset shows $\omega G_R$ versus $\log(H)$, exhibiting a QSMT around $\sim0.6$ Oe at 60 mK.}
	\label{fig4}
\end{figure} 

In Fig.~\ref{fig4}, we present the dramatic magnetic-field dependence of complex ac conductance at different temperatures, measured at 100 kHz. The inductive response $\omega G_I$ (red) is rapidly weakened in applied perpendicular magnetic field, as also found in amorphous InO$_x$ \cite{crane_survival_2007}, indicating a strong suppression of phase coherence among the superconducting grains. At low temperatures and $\approx 300$ Oe, $\omega G_I$ exhibits a knee feature, above which the superfluid density quickly drops to zero. Similar features were seen in previous MI measurements of InO$_x$ and MoGe thin films and attributed to a BKT-type transition, where the field-driven suppression of $T_\text{BKT}$ alluded an intervening metallic phase \cite{yazdani_observation_1993,misra_measurements_2013}. 

The dissipative response (blue), or the ac magneto-conductance $\omega G_R$, started off from zero at $H=0$ and the lowest temperature, pertaining to our assumption. As shown in the inset, within just $\sim 1$ Oe, we were able to detect the sharp onset of finite dissipation beyond our measurement sensitivity ($\sim 10^{-3}$ nH), in contrast to $G_R(\omega\neq0)=0$ in a fully gapped superconductor. The finite dissipation persists in the zero-resistance superconducting state until complex ac conductance diminishes at $\sim 500$ Oe.

\subsection{Complex ac conductance spectrum}

\begin{figure}[t]
	\centering
	\begin{subfigure}{}
		\includegraphics[width=\columnwidth]{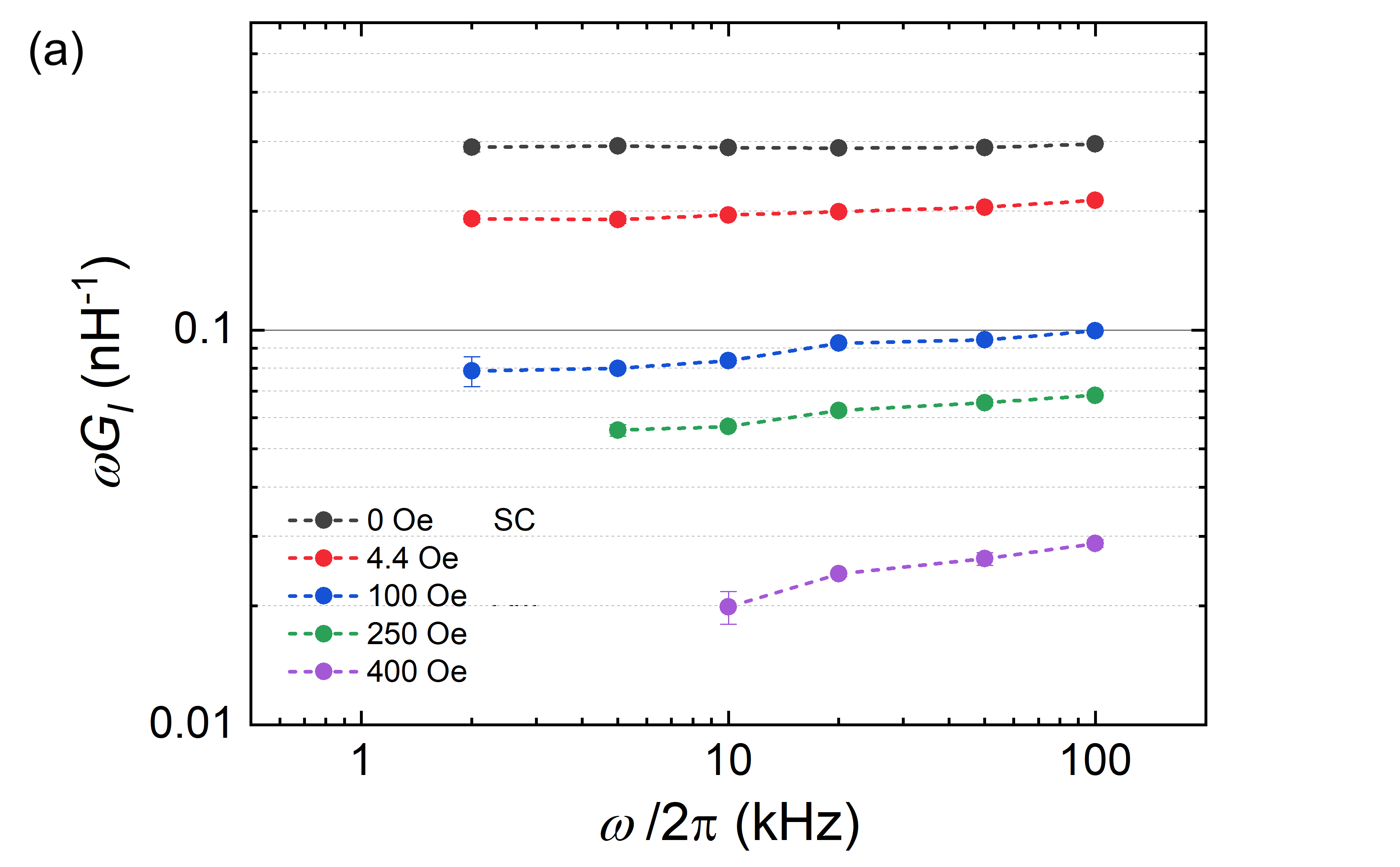}\vspace{-13.5mm}
	\end{subfigure}
	\begin{subfigure}{}
		\includegraphics[width=\columnwidth]{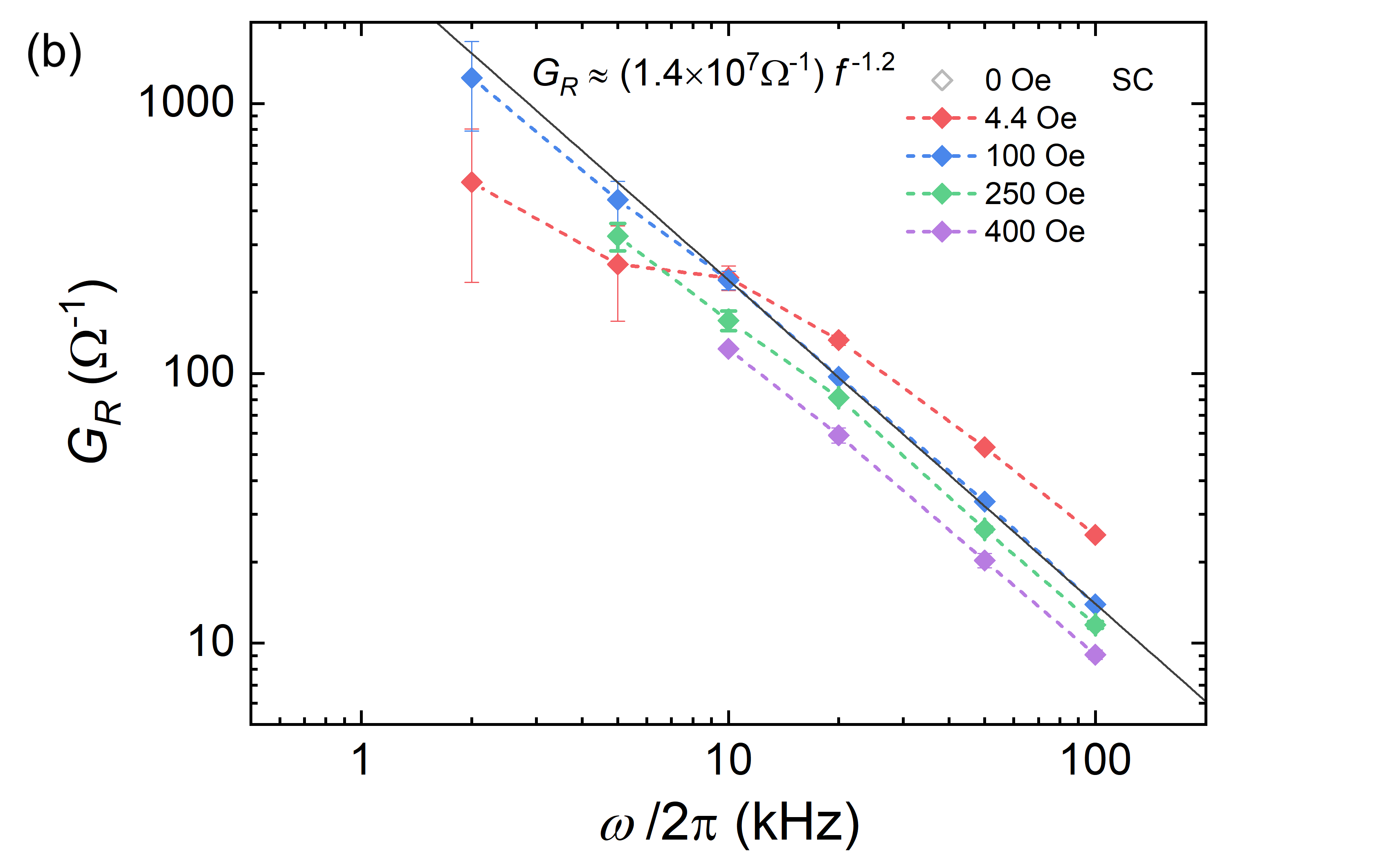}\vspace{-13.5mm}
	\end{subfigure}
	\begin{subfigure}{}
		\includegraphics[width=\columnwidth]{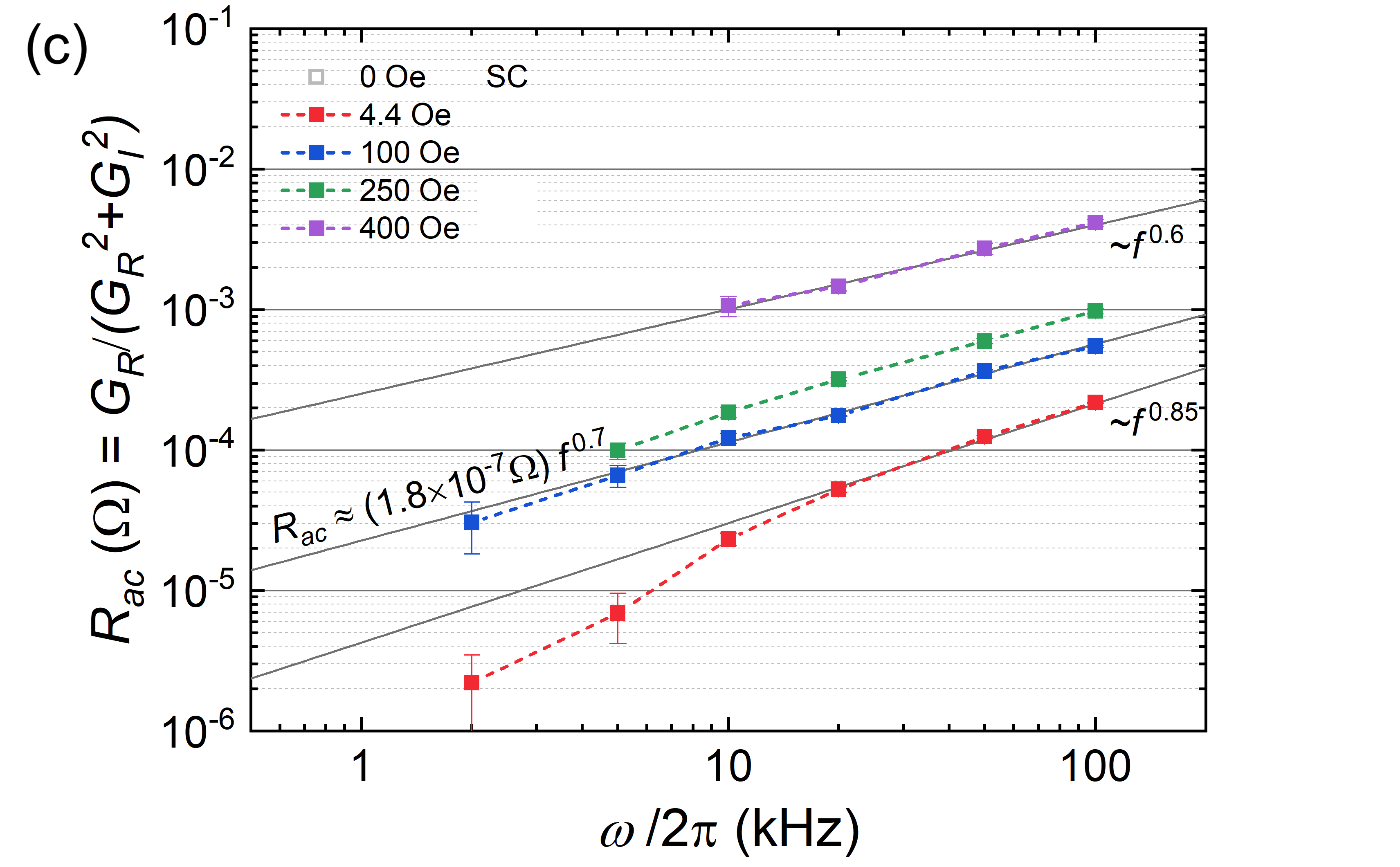}
	\end{subfigure}
	\caption{Complex ac conductance spectra at the lowest temperature $T\approx 60$ mK. \textbf{(a)} Inductive supercurrent response $\omega G_I$ shows weak positive frequency dependence. \textbf{(b)} Dissipative response $G_R$ exhibits anomalous power-law spectra in finite field. \textbf{(c)} Calculated ac resistance also depends on frequency as power laws in finite field.}
	\label{fig5}
\end{figure}

In this section, we discuss the frequency dependence of the measured complex ac response $\omega G_I$ and $G_R$ in a frequency range of 2-100 kHz, as shown in Fig.~\ref{fig5}. The imaginary part, \textit{i.e.} the inductive response of screening supercurrent, is frequency-independent at $H=0$ but acquires a weak positive correlation with frequency at finite $H$. The inductive response $\omega G_I$ is related to the superfluid density. For a higher frequency, the screening supercurrent spans a shorter distance that depends on the frequency as $L_\omega\sim\sqrt{D/\omega}$, where $D$ is a diffusion constant. In our granular system, an applied magnetic field would first suppress the weakest inter-grain Josephson coupling, creating regions of coupled grains that shrink as field increases \cite{zhang_anomalous_2022}. Consequently, higher frequency corresponds to screening current on a shorter length scale, which are generally more robust in larger magnetic field. Therefore, the $\omega G_I$ spectrum (Fig.~\ref{fig5}(a)) can be rationalized by invoking the effect of a broad granular distribution in our In/InO$_x$ system. 

The real part $G_R$ of complex ac conductance is expected to be zero in a fully gapped superconductor as $T\to0$. By assumption, we have fixed $G_R=0$ at $H=0$ to reverse the measurement circuit phase shift, and the zero-field $G_R$ is not plotted in Fig.~\ref{fig5}(b). However, in finite magnetic field, where resistance measurement indicated a zero-resistance superconducting ground state, $G_R$ shows anomalous frequency dependence in our further-annealed granular In/InO$_x$ system. Within 4.4-400 Oe (above 10 kHz for 4.4 Oe), the dissipative response exhibits power-law spectra $G_R\propto\omega^\alpha$ of approximately the same exponent $\alpha\approx -1.2$, where the increasing field only lowers the proportionality factor. We note similar observations in the superconducting state for previous microwave-frequency measurements of amorphous InO$_x$ thin film \cite{liu_microwave_2013,wang_absence_2018}, where an apparent anomalous metallic phase was found near a field-tuned quantum superconductor-metal transition. There, within the zero-resistance state, $G_R$ exhibited power-law spectra of similar exponents, while $G_I$ displayed spectra of positive slopes in magnetic field.

Furthermore, in Fig.~\ref{fig5}(c), we examine the real part of ac resistance calculated from the complex ac conductance as
\begin{equation}
	R_{ac}\equiv Re[\tilde{G}^{-1}]=\frac{G_R}{G_R^2+G_I^2}
\end{equation}
For a fully gapped superconductor as $T\to0$, both $G_{R,ac}$ and $R_{ac}$ are zero. In our sample, however, the ac resistance again follows an anomalous power law $R_{ac}\propto\omega^\beta$ in finite magnetic field. As field increases, the exponent $\beta$ decreases from 0.85 (above 10 kHz for 4.4 Oe) to 0.6 (400 Oe), whereas the magnitude increases pertaining to an expected positive magnetoresistance. The observed anomalous ac response, \textit{i.e.} power-law spectra for $G_R$ and $R_{ac}$, clearly suggest that the zero-resistance superconducting state in our granular system deviates from that of a fully gapped superconductor.

The power law spectra in ac response, together with an SIT in dc resistance, suggest that we may be probing the criticality near the superconductor-insulator quantum critical point at a finite yet very low frequency compared to the superconducting gap (note that $10\ \text{kHz}\sim5\ \mu\text{K}$). Extrapolating the $R_{ac}$ power-law spectrum to $\omega=0$ allows for three possibilities: (1) $R_{ac}$ drops to zero below a finite frequency indicating a true superconductor; (2) $R_{ac}$ saturates to a finite value as $\omega\to0$ indicating an anomalous metal; or (3) $R_{ac}$ remains a power law until $\omega=0$ at which $R_{ac}=0$ indicating an anomalous superconductor. In the last scenario, the system is a superconductor as measured by dc resistance, but the ac response is dissipative at any finite frequency.

\subsection{Comparison between ac conductance and ac resistance}

\begin{figure*}[t]
	\centering
	\begin{subfigure}{}
		\includegraphics[width=\columnwidth]{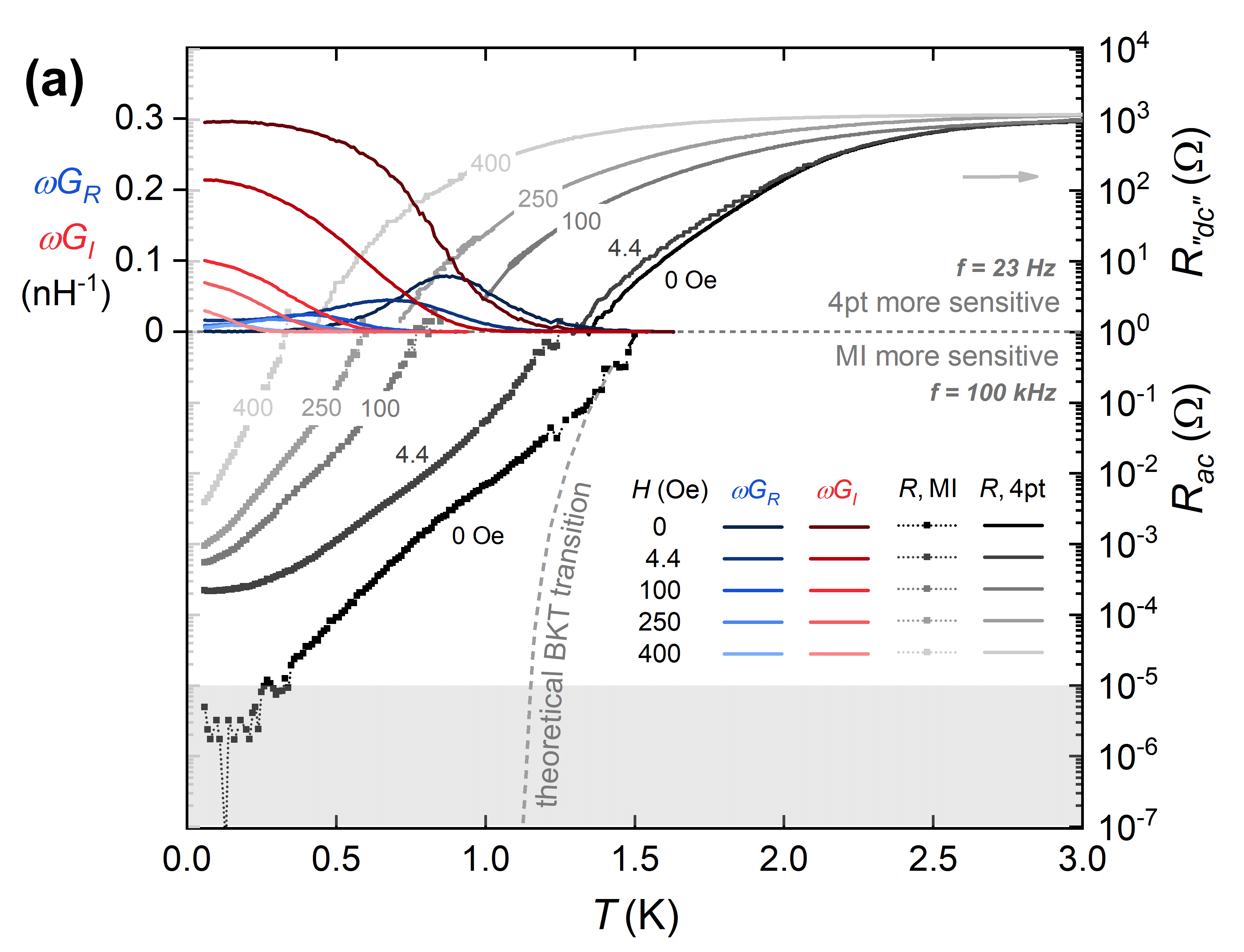}
	\end{subfigure}
	\begin{subfigure}{}
		\includegraphics[width=\columnwidth]{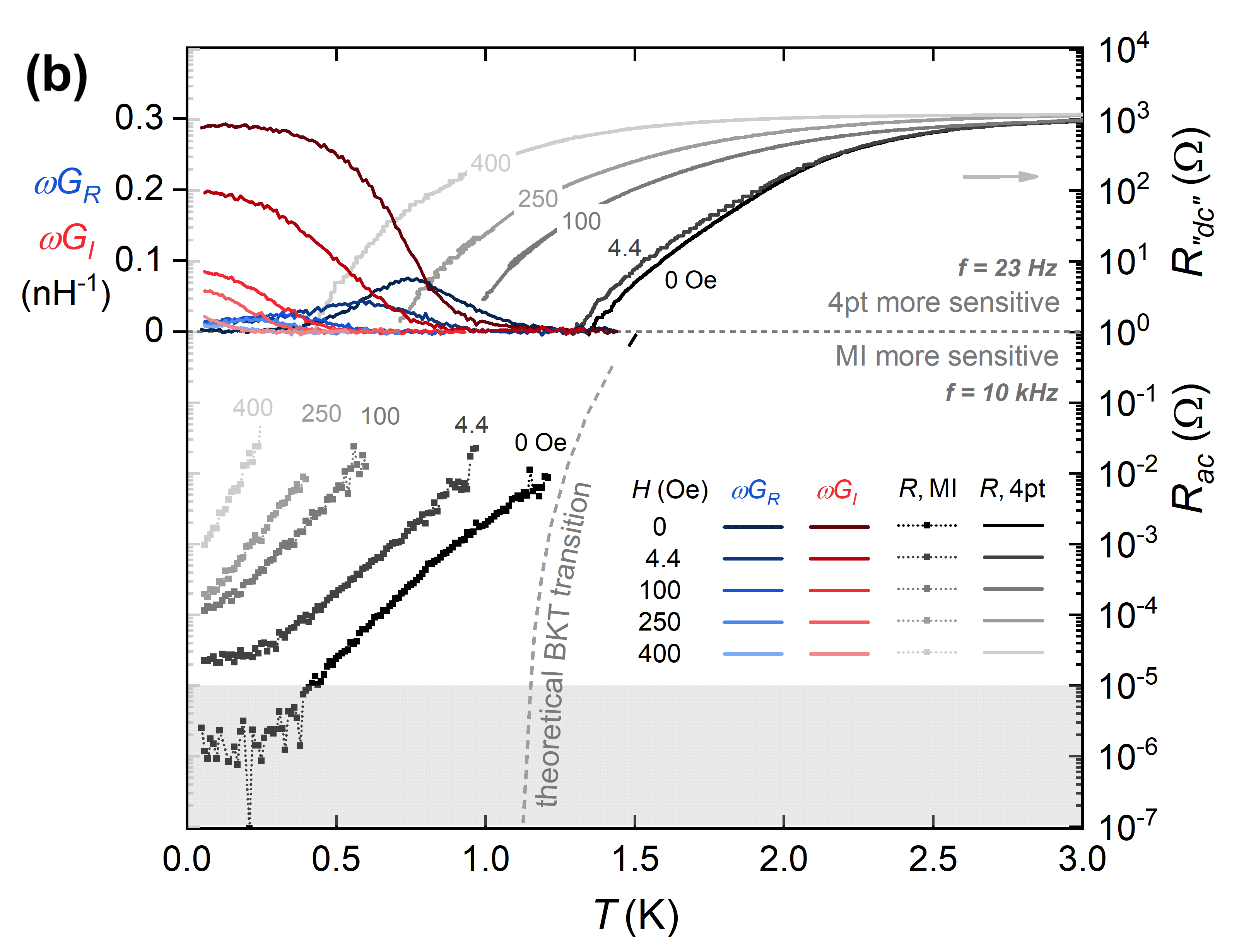}
	\end{subfigure}
	\caption{Comparison between four-point resistance, complex ac conductance, and ac resistance extracted from the MI measurements as a function of temperature at \textbf{(a)} 100 kHz and \textbf{(b)} 10 kHz. The left axes show the same data as in Figs.~\ref{fig3} and \ref{fig4}. The right axes consist of both resistance measurements, separated by $R\sim1\ \Omega$ above/below which 4pt/MI is more sensitive. See \textit{Supplemental Material} \cite{supplemental} for data processing details. The dashed lines represent a BKT transition as predicted by the Halperin-Nelson formula \cite{halperin_resistive_1979}. The shadow marks the sensitivity limit of resistance as converted from MI.}
	\label{fig6}
\end{figure*}

To better illustrate the evolution of the ac resistance (converted from complex ac conductance; referred to as ``MI'') as a function of temperature and magnetic field, in Fig.~\ref{fig6}, we plot it alongside the four-point resistance measurements (as in Fig.~\ref{fig1}; referred to as ``4pt''), overlaid with complex ac conductance data. 

Notwithstanding the stark difference between 4pt and MI measurements in terms of the measurement frequency, the  current vs. magnetic-field drive, and sample geometry \cite{supplemental}, the measured 4pt resistance and calculated ac resistance are in general agreement within the same order of magnitude. In Fig.~\ref{fig6} on the right axes, the 4pt measurements at 23 Hz, limited by a modest sensitivity $\gtrsim 1\ \Omega$, show an activated temperature dependence typically associated with thermally-excited unbounded vortices in 2d superconductors. Turning our focus to the MI measurements, we note that the nominal zero-field ac resistance, despite dropping below MI measurement sensitivity ($\sim10^{-5}\ \Omega$) as $T\to0$, deviates from the theoretical temperature dependence of a BKT transition \cite{halperin_resistive_1979}. The transition is significantly broadened, and the ac resistance depends exponentially on temperature $R_{ac}\sim\exp(T/T_0)$ spanning several orders of magnitude in $R$, where $T_0$ is the slope in $\ln R\text{ vs. }T$ plot.

Meanwhile, in a finite magnetic field, ac resistance saturates at a finite value as $T\to0$, consistent with the previous observation of finite dissipation with concomitant finite superfluid density. As shown in Fig.~\ref{fig6}(a), where the ac resistance is calculated from MI measurement at 100 kHz, the saturation value can be as low as $\sim2\times10^{-4}\ \Omega$ at 4.4 Oe, where the temperature dependence follows a similar exponential form. Similar behavior albeit in dc resistance was encountered in granular composites \cite{liu_resistive_1992,merchant_crossover_2001}, where the quantum tunneling of vortices or phase fluctuations were invoked to address the finite resistance as $T\to0$. We propose the same origin for the exponentially vanishing resistance here in granular In/InO$_x$, but employ the broad granular distribution to account for the anomalous temperature dependence \cite{zhang_robust_2021}. 

Turning to a lower frequency of 10 kHz, as in Fig.~\ref{fig6}(b), the magnitude of calculated ac resistance is reduced as indicated by the ac resistance spectrum discussed previously. The saturation also appears to be weakening, leading again to the question whether as $\omega\to0$, or in the dc resistance, there exists any saturation as $T\to0$. Despite the confirmation that superfluid density is robust in such a superconducting phase, the dc and ac conductivity may be dominated by gapless excitations as $T\to0$. The spectrum for these excitations determines whether the system will show an anomalous metallic state (presumably gapless, see Ref.~\cite{kapitulnik_colloquium_2019}) or an ``anomalous'' superconducting state with power-law spectra. Experimentally, gapless excitations in granular superconductors were previously  explored and related to strong phase fluctuations \cite{lamura_granularity-induced_2002}, whereas superconducting thin films of varying thickness were theoretically proposed as an example of gapless superconductors, along with superconductors with magnetic impurities and superconductor-metal interface under proximity effect \cite{de_gennes_superconductivity_1966}. Further theoretical and experimental efforts are required to establish the superconductor-metal-insulator transitions in granular superconductors as another embodiment of gapless superconductivity.

\section{Summary}

In summary, we systematically measured the full complex ac conductance of 2d granular In/InO$_x$ composites, using the MI technique. This system was previously shown to be a model system for a random array of resistively-shunted Josephson coupled grains can be tuned by annealing the InO$_x$ coupling layer to exhibit the full range of electronic phases dominated by Bose excitation. Focusing on the magnetic-field tuned ``true'' SIT, further annealing of the barrier material makes it more conductive, resulting in the emergence of intervening anomalous metallic states, originating from a ``failed superconductor'' at the low-field end to a ``failed insulator'' at the high field regime. The present paper completes this detailed study by further annealing the inter-grain barrier to the point where the anomalous metallic phase fades into the superconductor and the insulator weakens enough to lose its Bose-dominating character, revealing a new regime of superconductor-to-weak-insulator transition. This state is characterized by a unique complex ac conductance exhibiting coexistence of robust superfluid density and $T\to0$ saturating dissipation. In this regime, the complex conductance spectra featured an anomalous power-law frequency dependence, pointing to possible signatures of gapless superconductivity in such granular superconducting system. Since disordered superconductors are inherently granular in their superconducting state, often due to mesoscopic fluctuations in the disorder \cite{Skvortsov2005}, it is tempted to assign such behavior to other 2d disordered superconducting systems.

\bigskip

\noindent {\bf Acknowledgments:} We greatly benefited from discussions with Akshat Pandey and Steve Kivelson. Work at Stanford University was supported by the National Science Foundation Award Number: 2307132. Work at Tel-Aviv University was supported by the US-Israel Binational Science Foundation (Grant No. 2014098). 

\bibliography{InInOx_MI}

\onecolumngrid
\newpage
\setcounter{section}{0}
\setcounter{figure}{0}
\renewcommand{\thefigure}{S\arabic{figure}}
\renewcommand{\theequation}{S.\arabic{equation}}
\renewcommand{\thetable}{S\arabic{table}}
\renewcommand{\thesection}{S\arabic{section}}

\renewcommand{\thefootnote}{\fnsymbol{footnote}}

\begin{center}
	\textbf{SUPPLEMENTAL INFORMATION for} \\
	\vspace{1em}
	\textbf{Gapless superconductivity in the low-frequency electrodynamic response of two-dimensional granular In/InO$_x$ composites}\\
	
	\fontsize{9}{12}\selectfont
	
	\vspace{2em}
	Xinyang Zhang,$^{1,2}$ Jinze Wu,$^{1,3}$ Alexander Palevski,$^{4}$ and Aharon Kapitulnik,$^{1,2,3}$\\
	\vspace{1em}
	$^1${\it Geballe Laboratory for Advanced Materials, Stanford University, Stanford, CA 94305, USA}\\
	$^2${\it Department of Applied Physics, Stanford University, Stanford, CA 94305, USA}\\
	$^3${\it Department of Physics, Stanford University, Stanford, CA 94305, USA}\\
	$^4${\it School of Physics and Astronomy, Raymond and Beverly Sackler Faculty of Exact Sciences, Tel Aviv University, Tel Aviv 6997801, Israel}\\
\end{center}

\section*{List of supplemental contents:}
\begin{enumerate}[label=\Roman*.]
	\item Details of the mutual inductance coil experiment
	\item Performance characterization of the mutual inductance coil on a thick Nb film
\end{enumerate}

\newpage
\section*{I. Details of the mutual inductance coil experiment}
\begin{figure}[h]
	\centering
	\includegraphics[width=0.3\columnwidth]{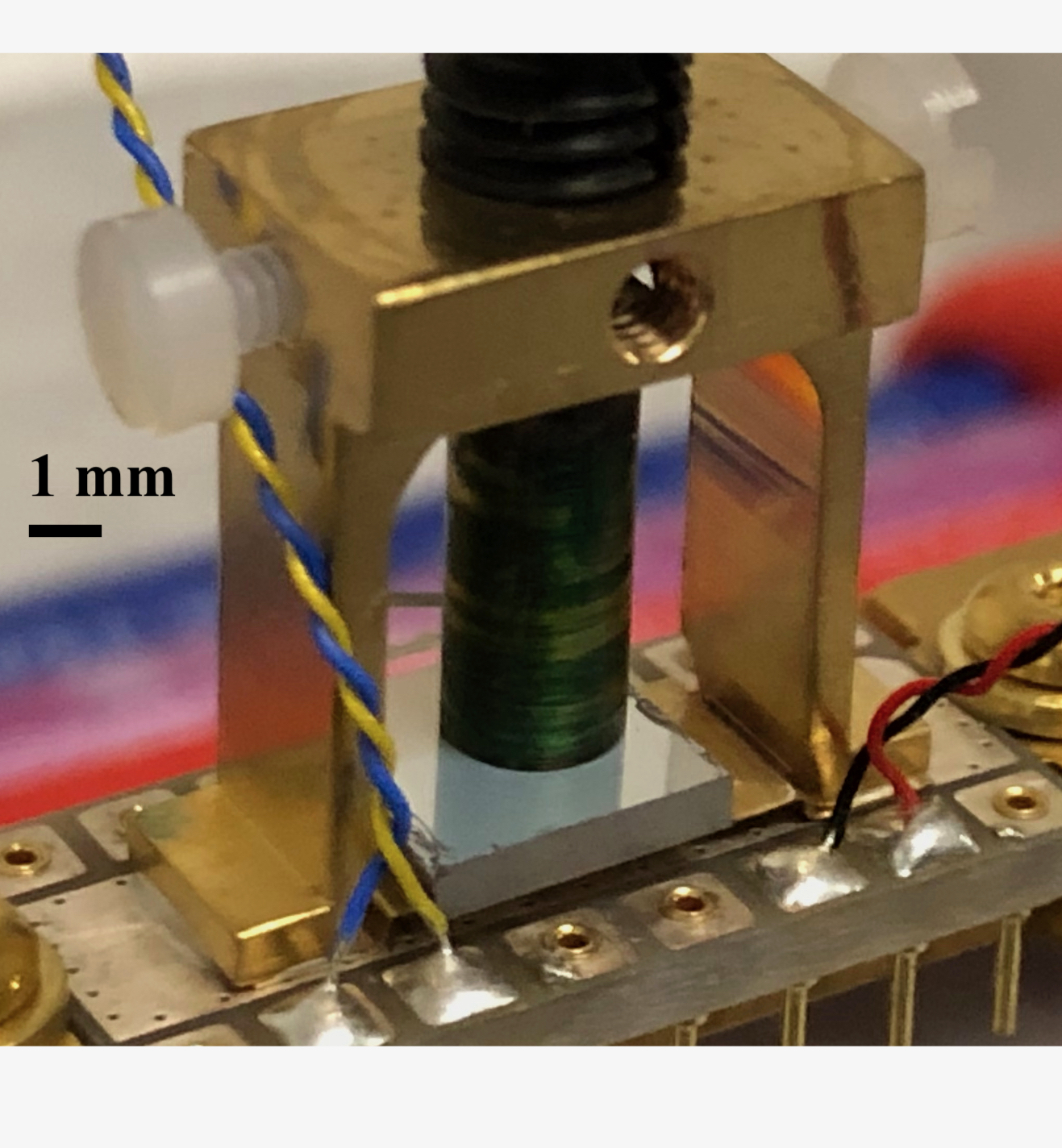}
	\includegraphics[width=0.4\columnwidth]{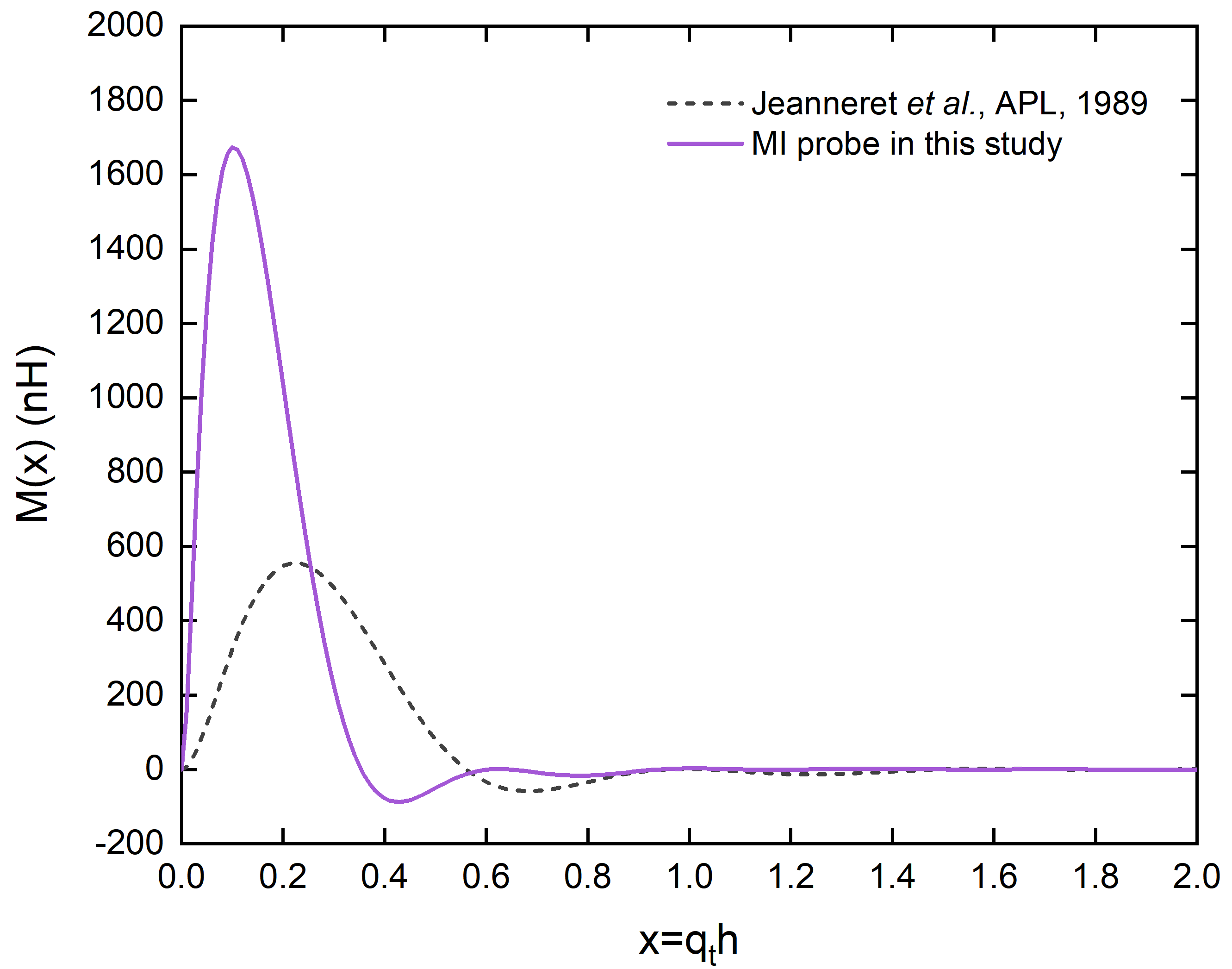}
	\includegraphics[width=\columnwidth]{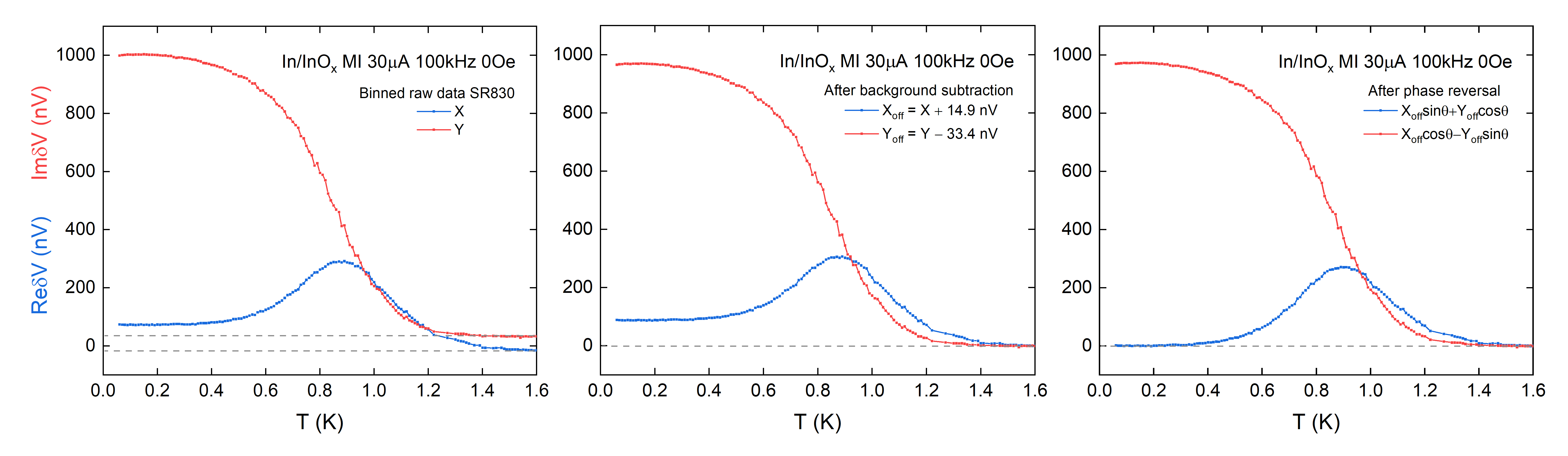}
	\caption{MI experiment and data analysis. (Upper left) MI probe used in this study under the microscope. The entire coil assembly was cast in epoxy and is mounted on a gold-plated copper mount. The platform where the sample is mounted on is a stamped gold-plated copper spring thermally anchored to a chip carrier. (Upper right) The mutual inductance function $M(x)$ for our MI probe as a function of dimensionless transverse spatial wave vector, explained in the supplementary text. Dashed line is $M(x)$ for the pioneering coil used in Jeanneret \textit{et al.} (Lower left) MI data processing procedure with an example data set in zero field at 100 kHz. The binned raw data are shown as directly measured in either quadrature of an SR830 lock-in amplifier. (Lower center) After the removal of the respective constant background, the two quadratures were set to zero $>1.5$ K. (Lower right) After a phase shift, the signal was decomposed into inductive and dissipative parts based on the assumption that $\text{Re}[V](H=0, T=0)=0$.}
	\label{figs1}
\end{figure}
Our mutual inductance (MI) probe is of the gradiometer-type consisting of compensated receive coils and a drive coil positioned co-axially with each other. The drive coil is supplied with an ac current of 30 $\mu$A, via a 100 k$\Omega$ current-limiting resistor. The resulting ac magnetic field is screened by the superconductor, which can be characterized by a complex ac conductance, producing a screening current in response. The magnetic field associated with this screening current in turn induces an emf in the receive coils, whose potential drop is measured by a lock-in amplifier. Assuming sample as an infinite plane [1], the induced voltage for a well-controlled coil geometry, which is captured by the mutual inductance function $M(x)$, is given by [2]:
\begin{equation}
	\delta V=i\omega I_D\int_0^\infty dx\ \frac{M(x)}{1+2x/i\mu_0 h\omega G}
	\label{S1}
\end{equation}

The geometrical parameters of our coils are: (receive) radius $r_R=0.75$ mm; wire diameter $\delta h_R=0.02$ mm; number of turns $N_R=56$; distance to sample $h_R\approx 0.02$ mm; (drive) radius $r_D=1.25$ mm; wire diameter $\delta h_D=0.08$ mm; number of turns $N_D=71$; distance to sample $h_D\approx 0.1$ mm. The coil assembly is based on a machined Nylon base, and was encapsulated in Stycast 1266 epoxy for mechanical integrity. The whole coil assembly is screwed into a gold-plated copper stand, which also serves as a good thermal anchor for the coil assembly. Electrical connections to the coils were made by silver epoxy. The sample was loaded on the cold finger into a thick layer of silver paint to reduce the stress exerted by the coil on the sample.

Fig.~\ref{figs1} shows the calculated mutual inductance function, \textit{i.e.} a geometrical factor that determines the sensitivity of the MI probe.
\begin{equation}
	M(x)=\pi\mu_0 h\alpha\beta J_1(\alpha x)J_1(\beta x)e^{-x}\frac{1-e^{N_D\gamma x}}{1-e^{-\gamma x}}\frac{1-e^{N_R\delta x}}{1-e^{-\delta x}}
	\label{S2}
\end{equation}
where $h=h_R+h_D$ is the sample-coil distance, the dimensionless $x=q_t h$ is the transverse spatial wave vector in units of $1/h$, and $\alpha,\beta,\gamma,\delta$ are $R_D,R_R,\delta h_D,\delta h_R$ in units of $h$, respectively. $J_1(x)$ is the $n=1$ Bessel function of the first kind. With our coil geometry, the calculated mutual inductance function generally behaves as a Bessel function, where the inductance starts out as zero at $x=0$, i.e. insensitive to spatially uniform magnetic flux. The largest inductance occurs at $x=1/4$, where $q_t=1/4h\sim 2.5$ mm$^{-1}$, corresponding to a wavelength roughly twice the diameter of our receive coil corresponding to the maximal flux. At $x=1/2$ the wave vector $q_t\sim 5$ mm$^{-1}$, corresponding to the diameter of our receive coil, so the flux equals zero. For longer wavelengths the signal is largely canceled and therefore we are mostly sensitive to magnetic fields produced by the screening current with a wave vector around $q_t=1/4h$.

In this study, the In/InO$_x$ film is several times larger than the receive coils, so Eqs.~\ref{S1} and \ref{S2} was directly applied for calculating the complex conductance. Nevertheless, due to the insufficient compensation between the astatic receive coil windings, or any stray capacitive coupling between the coaxial transmission lines for coil wiring, there exists a \textit{temperature-independent} background to either quadrature and a phase shift that mixes the two quadratures. In Fig.~\ref{figs1} we present the data processing procedure that converts the measured voltage to in-phase and out-of-phase responses of the sample. Here we assume that the dissipative response equals zero at $T=0$ and $H=0$, from which we extract a phase shift related to the capacitive couplings. We enforce the same phase shift for all measurements under a certain frequency.

\section*{II. Performance characterization of the mutual inductance probe on a thick niobium film}
We performed characterization measurements of the performance of our mutual inductance (MI) coil. Despite the careful construction of the coil assembly, imperfect cancellation between the drive and receive coils is inevitable, and background subtraction is required to extract voltage response from the sample. Furthermore, the measurement circuit, a complex RLC circuit itself, adds a phase shift to the measured voltage which needs to be compensated to extract the true real and imaginary parts of the voltage response. The goal of this characterization was to elucidate the current-frequency response of the coil, together with that of the measurement circuit. 

Here, we show the measured voltage response as a function of temperature, magnetic field, drive current, and drive frequency in a 5000-\AA\ Nb thick film. At the thickness of 500 nm, its superconducting transition temperature was found close to the bulk $T_c\approx 9.23$ K [3], and the penetration depth $\lambda(0)\sim 80$ nm [4]. As a type-II superconductor with $\kappa\sim 1$, the transition $\sim 10^{-2}\times T_c$ is fairly narrow and the lower critical field $H_{c1}\sim300$ Oe. Therefore, the zero-field transition as measured by MI should show a sharp onset of inductive response at $T_c$ and rapid saturation to the maximum response, accompanied by a narrow peak in the middle of the transition. The benefit of a narrow transition and temperature-independent response is to enable current-frequency sweeps at 4 K (superconducting) and 10 K (non-superconducting), yielding a meaningful background subtraction between the two sweeps.

The temperature-independent response is a result of Eq.~\ref{S1}, which, in the case of a purely imaginary conductance $\tilde{G}(T)=-iG_I(T)=L_K^{-1}(T)/i\omega$, reads
\begin{equation}
	\delta V=i\omega I_D\int_{0}^{\infty}dx\ \frac{M(x)}{1+2x/\mu_0 h L_K^{-1}(T)}
\end{equation}
that is also purely imaginary. For a narrow transition, the kinetic inductance $L_K=L_K(0)$, and the voltage response $\delta V=i\omega I_D \hat{\mathcal{I}}_M\{M(x)\}$, where $\hat{\mathcal{I}}_M\{M(x)\}$ is an integral of the geometrical MI function $M(x)$ independent of temperature, frequency, or current. Consequently, the measured voltage should be bilinear in both frequency and current.

\subsection*{Superconducting transition of a 5000-\AA\ Nb thick film}

\begin{figure}[ht]
	\centering
	\includegraphics[width=0.49\columnwidth]{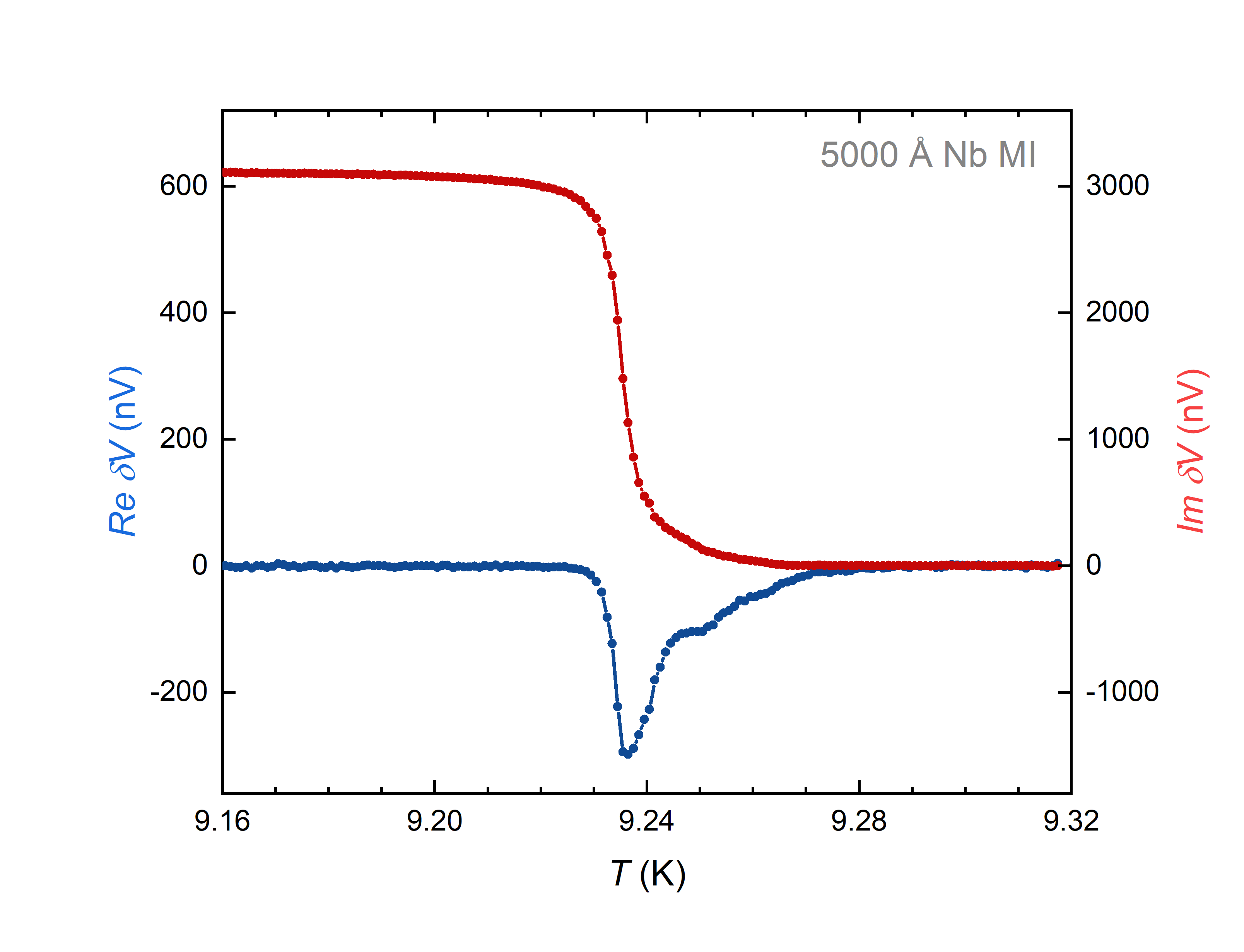}
	\includegraphics[width=0.49\columnwidth]{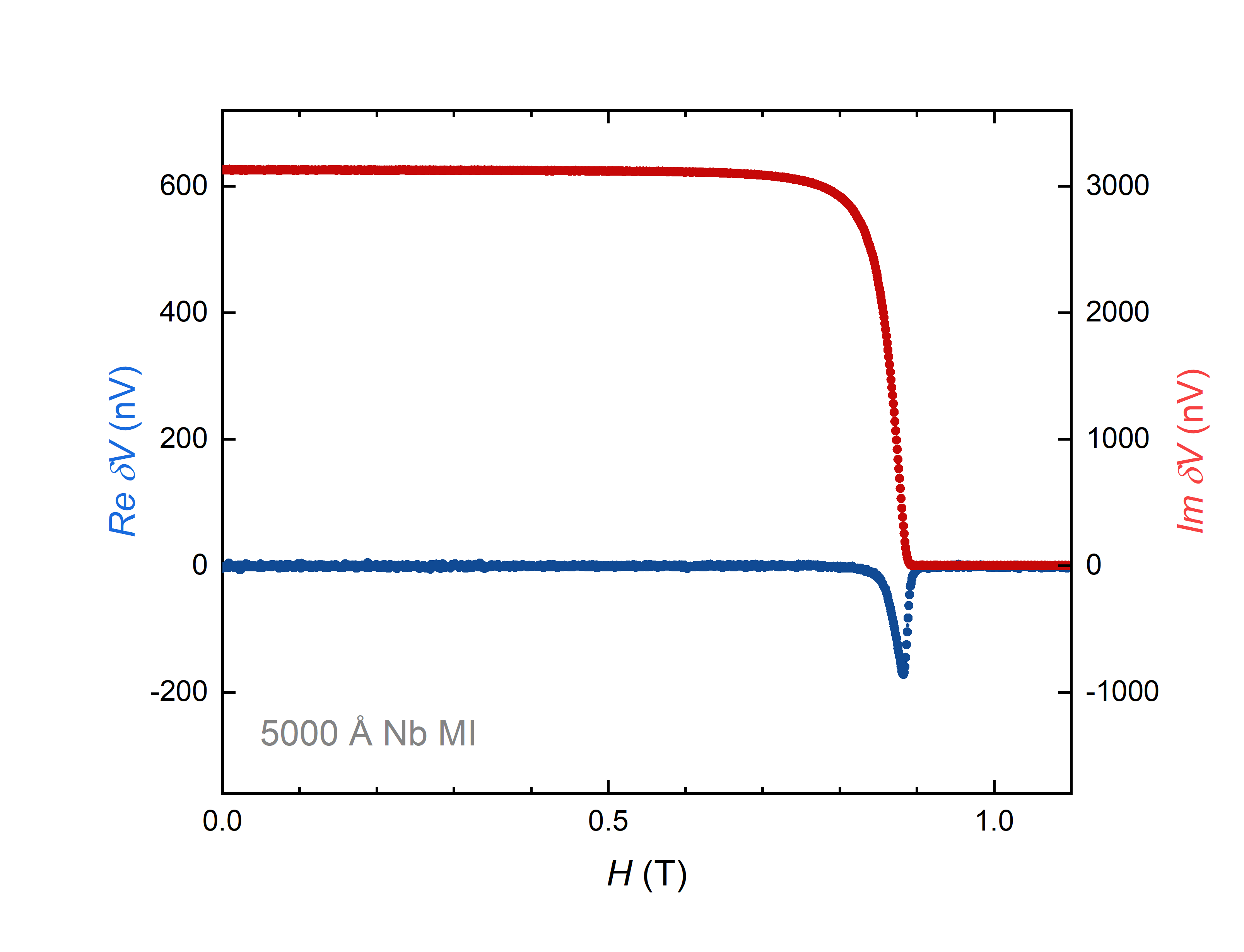}
	\caption[Voltage response of a thick Nb film from MI measurement]{Voltage response from MI measurement of the superconducting transition in 5000\AA\ Nb thick film. \textbf{(a)} Real (left) and imaginary (right) parts of the voltage response, indicating a transition temperature $T_c\approx 9.235$ K. \textbf{(b)} The voltage response as a function of perpendicular magnetic field, indicating an upper critical field $H_{c2}\approx0.88$ T.}
	\label{figs2}
\end{figure}

Fig.~\ref{figs2}(a) shows the real (dissipative) and imaginary (inductive) parts of voltage response measured across the receive coils. The inductive signal, corresponding to superfluid density $n_s\propto\lambda^{-2}$ at low fields, or equivalently kinetic inductance $L_K$ from the screening supercurrent, exhibits an onset at around 9.27 K. The signal gradually increases until 9.24 K, where the rapid superconducting transition takes place and the inductive response quickly saturates at the maximum value below 9.20 K. The dissipative signal, usually associated with vortex motion, creeps up slowly below 9.27 K, exhibiting a series of kink structures. At 9.235 K, it peaks at about 10\% of the inductive signal, before settling at zero, as expected in Meissner state.

Fig.~\ref{figs2}(b) show both channels of voltage response as a function of perpendicular magnetic field. At zero applied field $H=0$, the sample is in the Meissner state and the screening response is at the maximum. Above $H_{c1}\approx 0.1$ T, the inductive response begins to drop as vortices form an Abrikosov lattice. Although MI does not directly probe the bulk superconducting pair amplitude, whose suppression defines the upper critical field $H_{c2}$, the disappearance of inductive signal sets a lower bound to $H_{c2}\approx 0.88$ T. The dissipative part again exhibits a peak at the transition.

\subsection*{Current-frequency dependence of the measured inductive voltage signal}

\begin{figure}[ht]
	\centering
	\includegraphics[width=0.49\columnwidth]{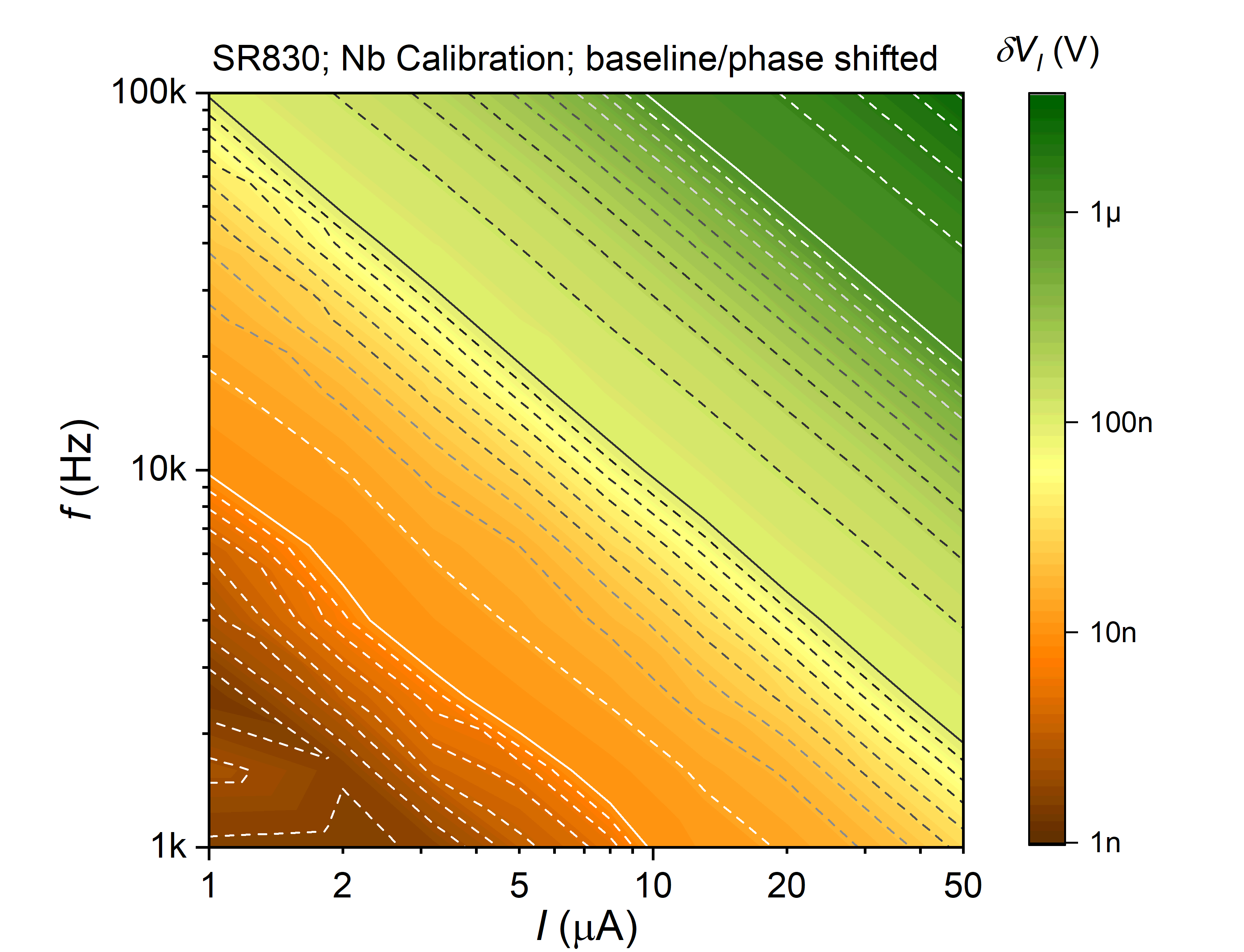}
	\includegraphics[width=0.49\columnwidth]{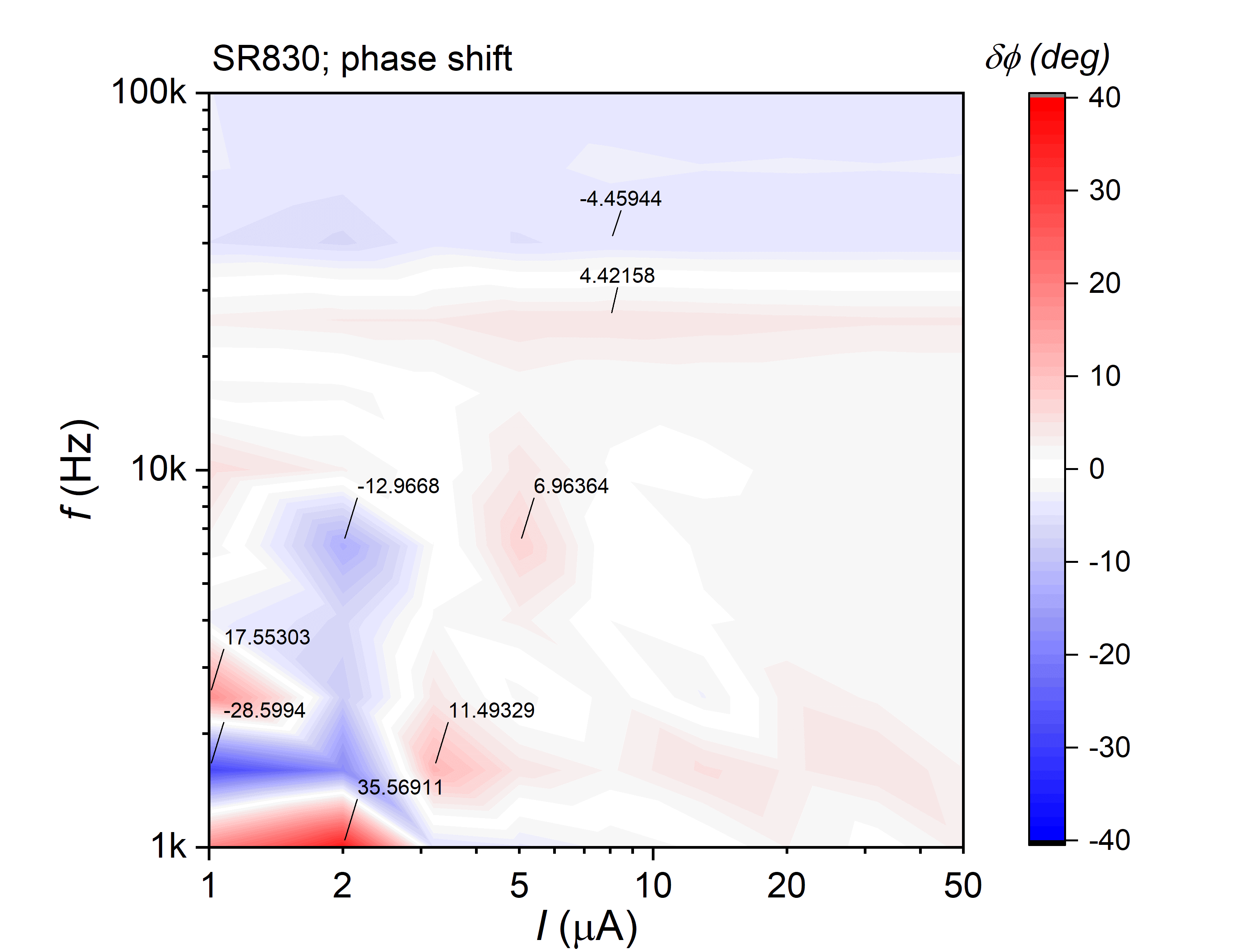}
	\caption{Voltage response of the Nb film versus drive current and frequency. \textbf{(a)} The inductive signal measured by SR830 lock-in amplifier. Each major level is traced by a solid line. The current-frequency mapping shows bilinear response. \textbf{(b)} Phase compensation as a function of drive current and frequency as measured by SR830 lock-in amplifier. Except for the lower left corner where phase noise is significant due to small signal, the phase shift is within $\pm5$ degrees.}
	\label{figs3}
\end{figure}

Fig.~\ref{figs3}(a) shows the magnitude of inductive response of the 5000-\AA\ Nb film as a function of drive frequency and current, after data processing. Current-frequency response sweeps were performed at both 4 K and 10 K, using Stanford research SR830 lock-in amplifier for $10\ \text{kHz}\leq f\leq100$ kHz. The full complex response at 4 K was subtracted by that of 10 K, which is presumed to be the measurement circuit background. The resultant ``sample'' response is then phase-compensated by reverting the phase shift, such that $\delta V_R(\text{4K})=0$, assuming zero dissipation in the sample at zero field and low temperatures. The response mapping as shown demonstrates a broad range of bilinear response in drive current $1\ \mu\text{A}\leq I\leq50\ \mu$A and frequency $1\ \text{kHz}\leq f\leq 100$ kHz, up to a maximum voltage of $\sim 3\ \mu$V and down to voltage measurement uncertainty of $\sim 2$ nV (with a time constant of 1 s). 

In finding $|\delta V_I|$ for Fig.~\ref{figs3}(a), a phase shift has been reversed for each frequency and current as needed to satisfy the $\delta V_R(T=0;H=0)$ assumption. Such phase compensations were plotted in Fig.~\ref{figs3}(b) for low-frequency region measured by SR830, where the low-frequency low-current corner corresponds to the lowest measured voltage both in the real and imaginary channel. There, the determination of phase compensation is contaminated by the uncertainty in voltage measurement of $\sim 2\ \text{nV}/\sqrt{\text{Hz}}$. Otherwise, the phase shift is generally within $\pm 5$ degrees, indicating minimal phase shift in the measurement circuit. 

\bigskip

\textbf{References}:\\

[1] S. J. Turneaure, A. A. Pesetski, and T. R. Lemberger, Numerical Modeling and Experimental Considerations for a Two-Coil Apparatus to Measure the Complex Conductivity of Superconducting Films, \textit{Journal of Applied Physics} \textbf{83}, 4334 (1998). \\

[2] B. Jeanneret, J. L. Gavilano, G. A. Racine, Ch. Leemann, and P. Martinoli, Inductive Conductance Measurements in Two‐dimensional Superconducting Systems, \textit{Appl. Phys. Lett.} \textbf{55}, 2336 (1989). \\

[3] B. T. Matthias, T. H. Geballe, and V. B. Compton, Superconductivity, \textit{Rev. Mod. Phys.} \textbf{35}, 1 (1963). \\

[1] A. I. Gubin, K. S. Il’in, S. A. Vitusevich, M. Siegel, and N. Klein, Dependence of Magnetic Penetration Depth on the Thickness of Superconducting Nb Thin Films, \textit{Phys. Rev. B} \textbf{72}, 064503 (2005).

\end{document}